\documentclass[reprint,superscriptaddress,prl,amsmath,amssymb,aps]{revtex4-2}

\usepackage{graphicx}% Include figure files
\usepackage{dcolumn}% Align table columns on decimal point
\usepackage{bm}% bold math
\usepackage{xcolor}
\usepackage{siunitx} 
\usepackage{relsize}
\usepackage{subfiles}
\usepackage{upgreek}
\usepackage{microtype} %for more homogeneous typesetting
\usepackage{listings}
\definecolor{codeframe}{RGB}{66,123,174}
\definecolor{codebackground}{RGB}{247,247,247}
\definecolor{mycolor}{RGB}{0, 114, 189}
\usepackage{newfloat}
\DeclareFloatingEnvironment[
    name=Supplementary Fig.,
    fileext=lof,
    within=section,
    placement=tbhp]{extendedfigure}
    
\usepackage[colorlinks=true, linkcolor=mycolor, citecolor=mycolor, urlcolor=mycolor, breaklinks=true]{hyperref}
\begin{document}

\title{Quantum-Nondemolition Probing of Multiphoton Transitions between Dressed States in a Strongly Driven Jaynes--Cummings System} 

\author{Han-Yu Ren}
\thanks{These authors contributed equally to this work.}
\affiliation{Hefei National Research Center for Physical Sciences at the Microscale and School of Physical Sciences, University of Science and Technology of China, Hefei 230026, China}
\affiliation{Shanghai Research Center for Quantum Science and CAS Center for Excellence in Quantum Information and Quantum Physics, University of Science and Technology of China,  Shanghai 201315, China}
\author{Zi-Yi Liu}
\thanks{These authors contributed equally to this work.}
\affiliation{Hefei National Research Center for Physical Sciences at the Microscale and School of Physical Sciences, University of Science and Technology of China, Hefei 230026, China}
\affiliation{Shanghai Research Center for Quantum Science and CAS Center for Excellence in Quantum Information and Quantum Physics, University of Science and Technology of China,  Shanghai 201315, China}
\author{Wen-Qi Fu}
\affiliation{Hefei National Research Center for Physical Sciences at the Microscale and School of Physical Sciences, University of Science and Technology of China, Hefei 230026, China}
\affiliation{Shanghai Research Center for Quantum Science and CAS Center for Excellence in Quantum Information and Quantum Physics, University of Science and Technology of China,  Shanghai 201315, China}
\author{Zhuo Chen}
\affiliation{Hefei National Research Center for Physical Sciences at the Microscale and School of Physical Sciences, University of Science and Technology of China, Hefei 230026, China}
\affiliation{Shanghai Research Center for Quantum Science and CAS Center for Excellence in Quantum Information and Quantum Physics, University of Science and Technology of China,  Shanghai 201315, China}
\author{Wen-Hua Qin}
\affiliation{Hefei National Research Center for Physical Sciences at the Microscale and School of Physical Sciences, University of Science and Technology of China, Hefei 230026, China}
\affiliation{Shanghai Research Center for Quantum Science and CAS Center for Excellence in Quantum Information and Quantum Physics, University of Science and Technology of China,  Shanghai 201315, China}
\author{Tao Rong}
\affiliation{Hefei National Research Center for Physical Sciences at the Microscale and School of Physical Sciences, University of Science and Technology of China, Hefei 230026, China}
\affiliation{Shanghai Research Center for Quantum Science and CAS Center for Excellence in Quantum Information and Quantum Physics, University of Science and Technology of China,  Shanghai 201315, China}
\author{Bo-Yang An}
\affiliation{Hefei National Research Center for Physical Sciences at the Microscale and School of Physical Sciences, University of Science and Technology of China, Hefei 230026, China}
\affiliation{Shanghai Research Center for Quantum Science and CAS Center for Excellence in Quantum Information and Quantum Physics, University of Science and Technology of China,  Shanghai 201315, China}
\author{ZhiAng Qin}
\affiliation{Henan Key Laboratory of Diamond Materials and Devices\mbox{,} Key Laboratory of Material Physics\mbox{,} Ministry of Education, School of Physics and Laboratory of Zhongxuan Light\mbox{,} Zhengzhou University\mbox{,} Daxue Road 75\mbox{,} Zhengzhou 450052\mbox{,} China}
\affiliation{Institute of Quantum Materials and Physics, Henan Academy of Sciences, Zhengzhou 450046, China}
\author{Li-Wei Liu}
\affiliation{Hefei National Research Center for Physical Sciences at the Microscale and School of Physical Sciences, University of Science and Technology of China, Hefei 230026, China}
\affiliation{Shanghai Research Center for Quantum Science and CAS Center for Excellence in Quantum Information and Quantum Physics, University of Science and Technology of China,  Shanghai 201315, China}
\author{Rami Mrad}
\affiliation{Hefei National Research Center for Physical Sciences at the Microscale and School of Physical Sciences, University of Science and Technology of China, Hefei 230026, China}
\affiliation{Shanghai Research Center for Quantum Science and CAS Center for Excellence in Quantum Information and Quantum Physics, University of Science and Technology of China,  Shanghai 201315, China}
\author{Ming Gong}
\affiliation{Hefei National Research Center for Physical Sciences at the Microscale and School of Physical Sciences, University of Science and Technology of China, Hefei 230026, China}
\affiliation{Shanghai Research Center for Quantum Science and CAS Center for Excellence in Quantum Information and Quantum Physics, University of Science and Technology of China,  Shanghai 201315, China}
\author{Yingqiu Mao}
\affiliation{Hefei National Research Center for Physical Sciences at the Microscale and School of Physical Sciences, University of Science and Technology of China, Hefei 230026, China}
\affiliation{Shanghai Research Center for Quantum Science and CAS Center for Excellence in Quantum Information and Quantum Physics, University of Science and Technology of China,  Shanghai 201315, China}

\author{Yuan Zhang}
\email{yzhuaudipc@zzu.edu.cn}
\affiliation{Henan Key Laboratory of Diamond Materials and Devices\mbox{,} Key Laboratory of Material Physics\mbox{,} Ministry of Education, School of Physics and Laboratory of Zhongxuan Light\mbox{,} Zhengzhou University\mbox{,} Daxue Road 75\mbox{,} Zhengzhou 450052\mbox{,} China}
\affiliation{Institute of Quantum Materials and Physics, Henan Academy of Sciences, Zhengzhou 450046, China}

\author{Johannes Majer}
\email{jmajer@ustc.edu.cn}
\affiliation{Hefei National Research Center for Physical Sciences at the Microscale and School of Physical Sciences, University of Science and Technology of China, Hefei 230026, China}
\affiliation{Shanghai Research Center for Quantum Science and CAS Center for Excellence in Quantum Information and Quantum Physics, University of Science and Technology of China,  Shanghai 201315, China}
\date{\today}

\begin{abstract}
The response of the Jaynes--Cummings (JC) model to strong driving has been examined previously, but the origin of intriguing features near emitter--cavity resonance remains poorly understood because of the large background in conventional transmission spectroscopy. Here we introduce a spectroscopic approach for circuit quantum electrodynamics that combines strong cavity driving with quantum-nondemolition qubit readout. Separating the drive and readout substantially enhances the measurement sensitivity as compared with conventional cavity-transmission spectroscopy. With this technique, we identify the near-resonant features as multiphoton transitions between emitter--photon dressed states of the same parity, and resolve transitions between dressed states of opposite parity. Away from resonance, we also observe selective enhancement and suppression of multiphoton transitions, together with transitions involving the second excited state of the anharmonic qubit. Our approach opens a route to resolving the rich transition structure of the JC ladder and probing strongly driven light--matter interactions.
\end{abstract}

\maketitle

The Jaynes--Cummings (JC) model is a fundamental model derived from quantum electrodynamics (QED) to describe light--matter interaction. Specifically, it describes the coherent coupling of a two-level system to a single electromagnetic mode, and predicts an anharmonic ladder of emitter--photon dressed states \cite{wallsQuantumOptics,larsonJaynesCummingsModel2024}. Studies based on the JC model have revealed a rich variety of quantum phenomena, such as Mollow-like multiplets in quantum dots \cite{kasprzakJaynesCummingsLadder2010, moelbjergResonanceFluorescenceSemiconductor2012, allcockQuantumMollowQuadruplet2022}, phonon blockade in trapped ions \cite{toyodaExperimentalRealizationQuantum2013, debnathObservationHoppingBlockade2018}, and $\sqrt{n}$-nonlinear splittings in Rydberg atoms \cite{leeDemonstrationJaynesCummingsLadder2017}. Meanwhile, circuit quantum electrodynamics (circuit-QED), in which a superconducting qubit couples to a microwave cavity, is a versatile platform for direct spectroscopic exploration of the JC model \cite{blaisCavityQuantumElectrodynamics2004,wallraffStrongCouplingSingle2004}, including vacuum Rabi splitting, multiphoton transitions from the ground state to higher dressed states, photon-number-dependent transitions \cite{schusterResolvingPhotonNumber2007,finkClimbingJaynesCummings2008a,bishopNonlinearResponseVacuum2009}, and thermally assisted transitions \cite{finkThermalExcitationMultiphoton2009,finkQuantumToClassicalTransitionCavity2010,bonsenProbingJaynesCummingsLadder2023}.
These phenomena are commonly studied using conventional transmission or reflection spectroscopy, in which the cavity is driven by a microwave field and its transmitted or reflected signal is measured as a function of microwave frequency.

Previous circuit-QED studies reported spectral features near qubit--cavity resonance and attributed them to bistability \cite{dombiBistabilityEffectExtreme}, photon-blockade breakdown, and critical slowing down \cite{finkObservationPhotonBlockadeBreakdown2017,brookesCriticalSlowingCircuit2021,settEmergentMacroscopicBistability2024}. A simple analysis of the dressed-state ladder of the JC model, however, also predicts multiphoton transitions originating from dressed states at similar drive frequencies. These transitions have been difficult to resolve because strong driving substantially broadens the spectrum. 
Two effects contribute to this broadening: First, the probe and drive fields share the same cavity, which produces a strong background signal. Second, the strong drive required to populate excited dressed states further dresses the JC ladder through drive-induced state mixing, resulting in the supersplitting observed in conventional transmission spectroscopy \cite{bishopNonlinearResponseVacuum2009}.
To identify the origin of the features near qubit--cavity resonance, we mitigate spectral broadening by reading out the qubit through a quantum-nondemolition (QND) measurement while driving the cavity independently. Although QND qubit readout is widely used in quantum computing \cite{blaisCavityQuantumElectrodynamics2004,lupascuQuantumNondemolitionMeasurement2007}, to our knowledge it has not been applied to probe the allowed transitions of the JC model. 
\begin{figure}[bp]
 \includegraphics[width=0.83\columnwidth]{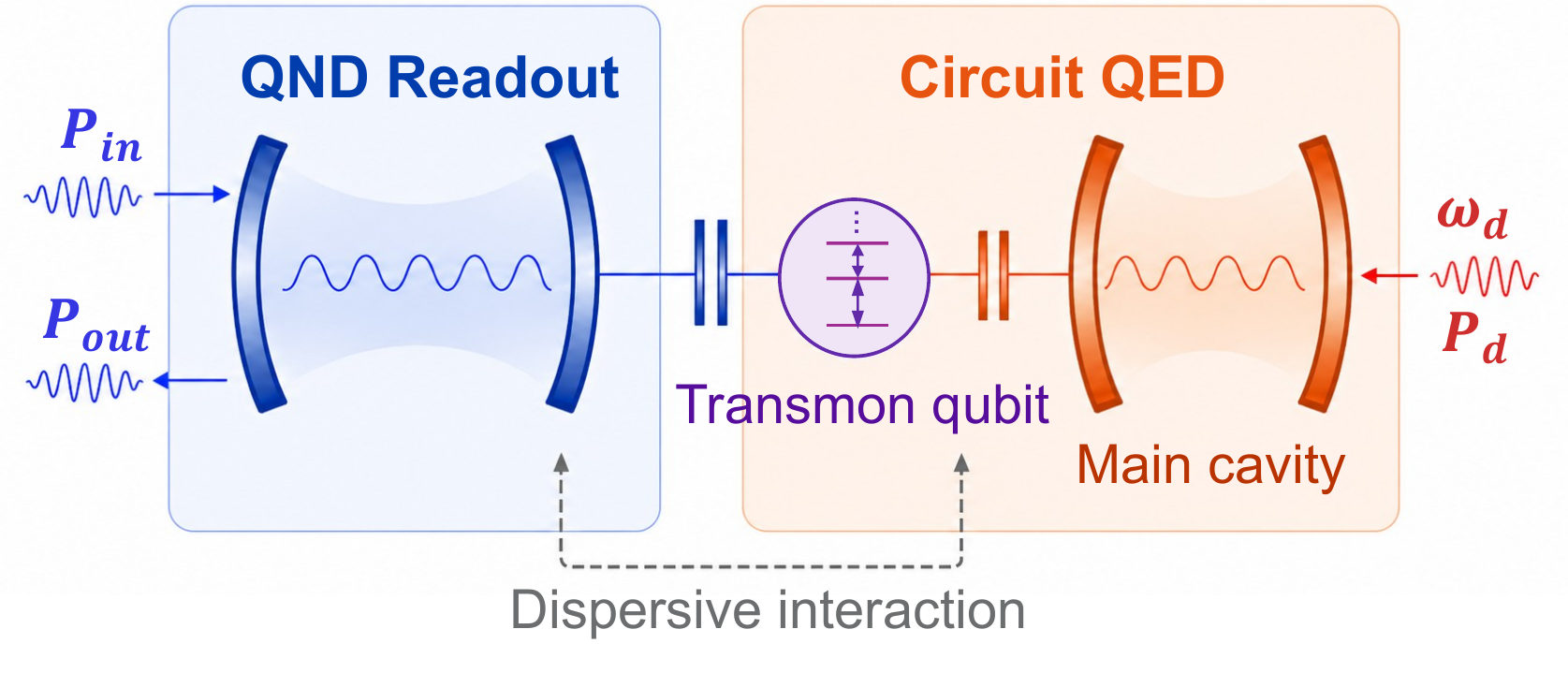}    \caption{\label{fig:fig1}\textbf{Simplified schematic of the circuit-QED system and its quantum-nondemolition (QND) readout configuration.} A strong microwave drive is applied to the cavity--qubit system (right), while the transmon occupation is measured through the independent readout channel (left), separating the drive and readout. The inline Purcell filter is omitted for clarity; experimentally, we measure the transmission $S_{21}$ of the complete readout channel.}
\end{figure}
\begin{figure*}[tbp]
    \includegraphics[width=0.98\textwidth]{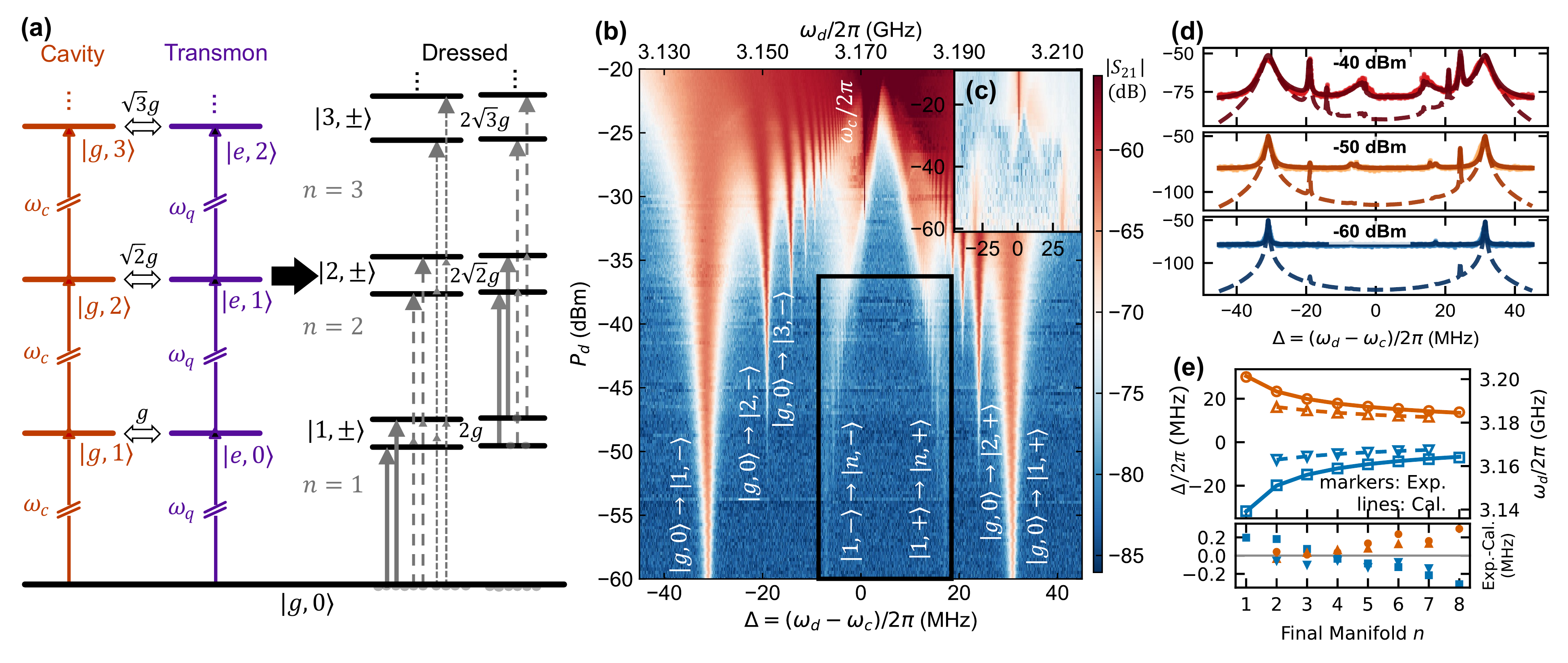}   \caption{\label{fig:fig2}\textbf{QND-based spectroscopy near qubit--cavity resonance.}
    \textbf{(a)} Energy-level diagram of the Jaynes--Cummings model. The left side shows the coupling between qubit--photon product states, while the right side shows the resulting dressed states and the single-, two-, and three-photon transitions (solid, dashed, and dotted arrows, respectively).
    \textbf{(b)} Measured readout-channel transmission $S_{21}$ as a function of drive detuning $\Delta=(\omega_d-\omega_c)/2\pi$ and drive power $P_d$. Two distinct sets of spectral features are visible: bright outer transition sets at larger $|\Delta|$, arising from conventional multiphoton transitions starting from $\lvert g,0\rangle$, and narrow inner transition sets within the black box starting from $|1, \pm\rangle$. At high drive powers, the inner transition sets approach a small positive detuning from the bare-cavity frequency. The inset \textbf{(c)} shows the conventional cavity-transmission spectrum measured over the same drive-frequency and drive-power window; the thermally assisted transitions are largely obscured even after much longer averaging \cite{SuppMat}.
    \textbf{(d)} Comparison of experimental line cuts at selected weak and intermediate drive powers with steady-state master-equation calculations for a finite cavity occupation $n_{\rm th}=0.010$ (solid curves) and vanishing occupation $n_{\rm th}=0$ (dashed curves).
    \textbf{(e)} Comparison of the manually selected transition frequencies (symbols) with the calculated values (lines). The lower panels show the residuals $(\omega_{\mathrm{exp}}-\omega_{\mathrm{cal}})/2\pi$. A small residual qubit--cavity detuning, $\Delta_{qc}=(\omega_q-\omega_c)/2\pi\approx0.47~\mathrm{MHz}$, is included to obtain agreement between theory and experiment.}
\end{figure*}
With this technique, we resolve the aforementioned features as thermally assisted multiphoton transitions between dressed states of the same parity, and observe weaker transitions between states of opposite parity, which are strongly suppressed in the idealized parity-selection picture. Furthermore, by tuning the qubit frequency, we observe selective enhancement and suppression of these transitions due to changes in the photonic and qubit contributions to the dressed states, as well as transitions involving the second excited state of the anharmonic qubit. Thus, our study resolves the rich transition structure of the JC ladder and opens a route for probing strongly driven light--matter interactions.

Our system is schematically shown in Fig.~\ref{fig:fig1}. A strong spectroscopy tone drives the cavity--transmon subsystem, while an independent readout resonator dispersively coupled to the transmon enables QND detection. The simplified diagram omits the inline Purcell filter~\cite{jeffreyFastAccurateState2014}; in the physical device, the measured readout signal is the transmission $S_{21}$ of the complete Purcell-filter channel. By separating the drive and readout, this scheme substantially reduces the driven-cavity background and preserves weak transitions that remain resolvable. Further details are available in the Supplementary Material (SM) \cite{SuppMat}.

Theoretically, the dynamics of the circuit-QED system can be described by the master equation
\begin{align}
\frac{\partial \hat{\rho}}{\partial t}
=&-\frac{i}{\hbar}[\hat{H}_0+\hat{H}_d,\hat{\rho}]\nonumber\\
&+\kappa(n_{\rm th}+1)\mathcal{D}[\hat{a}]\hat{\rho}
+\kappa n_{\rm th}\mathcal{D}[\hat{a}^\dagger]\hat{\rho}
\nonumber\\
&+\gamma_1\mathcal{D}[\hat{b}]\hat{\rho}
+\gamma_\varphi\mathcal{D}[\hat{b}^\dagger \hat{b}]\hat{\rho},
\label{eq:master}
\end{align}
In this equation, the first line describes the coherent dynamics. The Hamiltonian reads
\begin{equation}
\frac{\hat{H}_0}{\hbar}
=
\omega_c \hat{a}^\dagger \hat{a}
+
\omega_q \hat{b}^\dagger \hat{b}
-
\frac{\alpha}{2} \hat{b}^\dagger \hat{b}^\dagger \hat{b} \hat{b}
+
g\left(\hat{a}^\dagger \hat{b} + \hat{a} \hat{b}^\dagger\right),
\label{eq:H0}
\end{equation}
where $\hat{a}$ and $\hat{b}$ denote the annihilation operators for the cavity and transmon modes, respectively. The bare cavity frequency is $\omega_c/2\pi=3.171~\mathrm{GHz}$, whereas the flux-tunable bare qubit frequency spans the range $\omega_q/2\pi=2.800\text{--}3.856~\mathrm{GHz}$. The cavity--qubit coupling strength is $g/2\pi=31~\mathrm{MHz}$, and $\alpha/2\pi=192~\mathrm{MHz}$ is the transmon anharmonicity. For the resonant measurements, $\omega_q$ is tuned nominally into resonance with $\omega_c$, with a small residual detuning $\Delta_{qc}=(\omega_q-\omega_c)/2\pi$, which arises from finite experimental tuning accuracy and is retained in the quantitative calculations.
The Hamiltonian
\begin{equation}
\frac{\hat{H}_d}{\hbar}
=
\epsilon_d
\left(
\hat{a} e^{i\omega_d t}
+
\hat{a}^\dagger e^{-i\omega_d t}
\right),
\label{eq:Hd}
\end{equation}
describes the microwave driving of the cavity. Here, $\omega_d$ is the the drive frequency and $\epsilon_d$ is the drive amplitude.

% ($1/T_2^* = 1/(2T_1)+\gamma_\varphi/2$)
The dissipative terms in Eq.~\eqref{eq:master} describe cavity loss as well as transmon relaxation and dephasing, with $\kappa/2\pi \approx 67~\mathrm{kHz}$, $\gamma_1/2\pi\approx 4.2~\mathrm{kHz}$ and $\gamma_\varphi/2\pi \approx 65~\mathrm{kHz}$. $\mathcal{D}[\hat{o}]\hat{\rho}=\hat{o}\hat{\rho}\hat{o}^\dagger-\{\hat{o}^\dagger\hat{o},\hat{\rho}\}/2$ is the Lindblad dissipator for an operator $\hat{o}$. We parameterize the effective mean cavity occupation using the finite-temperature Bose--Einstein distribution,
$n_{\rm th}=[\exp(\hbar\omega_c/k_{\rm B}T_{\rm eff})-1]^{-1}$. The refrigerator temperature, $T_{\rm fridge}=13~\mathrm{mK}$, provides an initial estimate of $n_{\rm th}$. We then determine $n_{\rm th}$ by treating it as a fitting parameter for spectra at different drive powers. The best-fit occupation is $n_{\rm th}=0.010$, indicating that the effective temperature of our system is $T_{\rm eff}\simeq33~\mathrm{mK}$. This effective temperature is comparable to, if not lower than, previously reported values for superconducting qubits operated at similar mixing-chamber temperatures~\cite{konoMechanicallyInducedCorrelated2024, jinThermalResidualExcitedState2015}.

We solve the master equation in the steady state using a density-matrix method and then calculate the mean transmon occupation $\langle\hat{b}^\dagger\hat{b}\rangle$ as a function of drive frequency and power. Through the dispersive coupling between the transmon and the readout resonator, this occupation is mapped onto the transmitted probe signal $S_{21}$ at a fixed frequency.

We first study the system near qubit--cavity resonance, $\omega_q\simeq\omega_c$. In this regime, the qubit can be modeled as a two-level system because the second excited transmon state is detuned from the two-excitation subspace by the anharmonicity. The corresponding JC Hamiltonian is
\begin{equation}
\frac{\hat H_{\rm JC}}{\hbar}=\omega_c\hat a^\dagger\hat a+\omega_q\hat\sigma_+\hat\sigma_-+g\left(\hat a^\dagger\hat\sigma_-+\hat a\hat\sigma_+\right),
\end{equation} 
where $\hat\sigma_+$ and $\hat\sigma_-$ are the raising and lowering operators of the two-level system. To proceed, we define the product states $\lvert a,n\rangle=\lvert a\rangle\lvert n\rangle$, where $a=g,e$ labels the qubit state and $n=0,1,\ldots$ is the photon number, as shown on the left side of Fig.~\ref{fig:fig2}(a). In the $n$-excitation subspace, $\lvert g,n\rangle$ couples to $\lvert e,n-1\rangle$ with strength $\sqrt{n}g$. At exact resonance, diagonalizing this two-state block gives the dressed states $\lvert n,\pm\rangle=(\lvert g,n\rangle\pm\lvert e,n-1\rangle)/\sqrt{2}$, separated in frequency by $2\sqrt{n}g$, as shown on the right side of Fig.~\ref{fig:fig2}(a). The drive Hamiltonian [Eq.~\eqref{eq:Hd}] therefore induces one-, two-, and three-photon transitions from $\lvert g,0\rangle$ to $\lvert n,\pm\rangle$ at frequencies $\omega_c\pm\sqrt{n}g/n$, indicated by the solid, dashed, and dotted arrows, respectively. If the first excited dressed states $\lvert 1,\pm\rangle$ have finite thermal populations, transitions can also occur from these states to higher dressed states $\lvert n,\pm\rangle$. Analogous transitions can also originate from higher thermally populated states such as $\lvert 2,\pm\rangle$; their visibility depends on the population distribution among the ground and dressed states.

Although the second excited transmon state $\lvert f\rangle$ does not qualitatively alter this transition structure, its off-resonant coupling produces dispersive shifts of the dressed-state energies. Consequently, at large $n$, the multiphoton-transition frequencies to both $\lvert n,-\rangle$ and $\lvert n,+\rangle$ lie above the bare-cavity frequency rather than converging to it as in an ideal two-level model. We therefore retain the full multilevel model for quantitative calculations and use the standard JC model to analyze the main features.

\begin{figure}[htbp]
 \includegraphics[width=0.85\columnwidth]{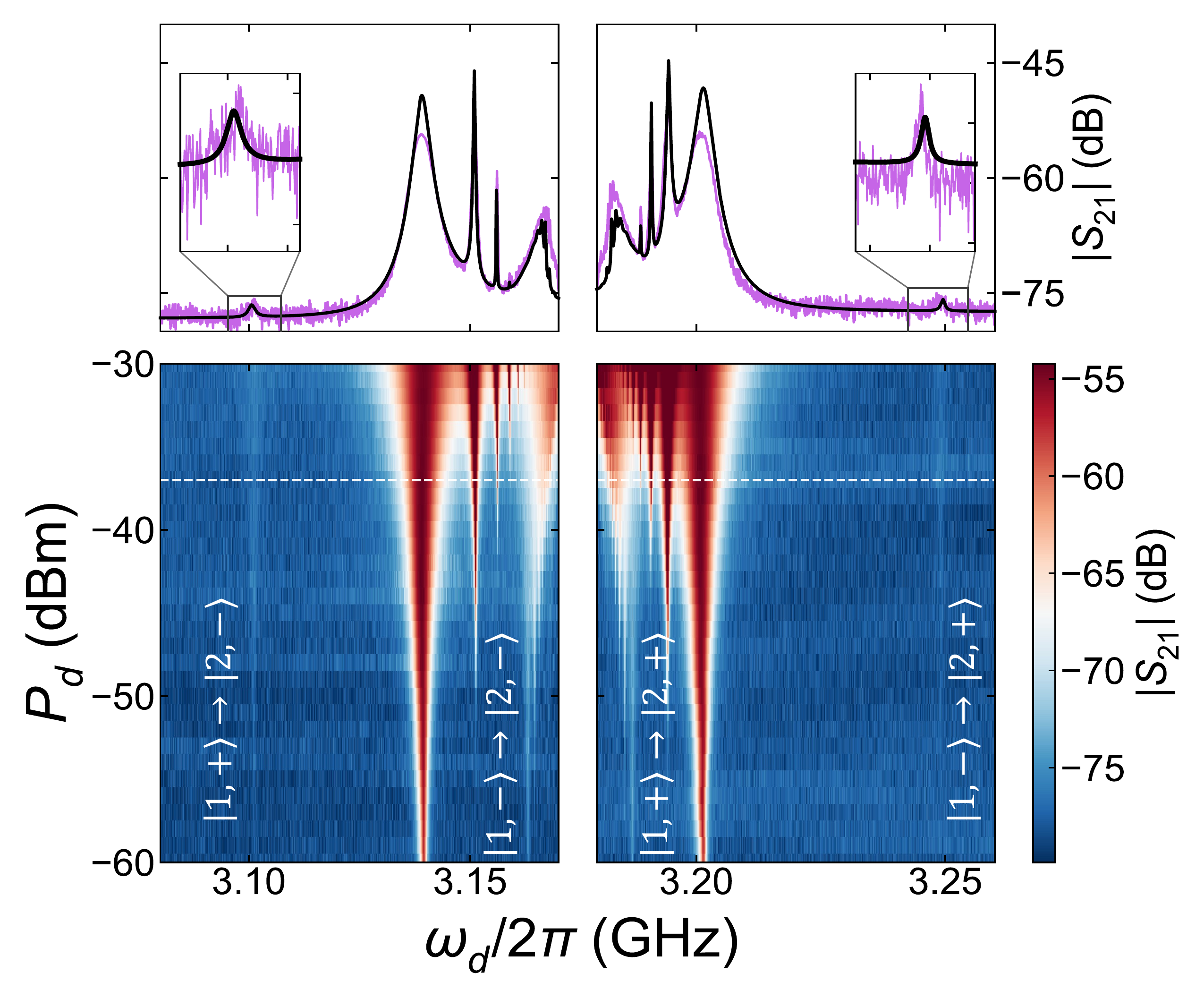}
    \caption{\label{fig:fig3}\textbf{QND probing of opposite-parity transitions.}
    The same transmission spectrum as shown in Fig.~\ref{fig:fig2}(b) is presented over a wider range of drive detuning. Faint features on the negative- and positive-detuning sides are attributed to the transitions $\lvert 1,+\rangle\to\lvert 2,-\rangle$ and $\lvert 1,-\rangle\to\lvert 2,+\rangle$ between dressed states of opposite parity, respectively. The upper panel compares the experimental line cut at a drive power of $-37~\mathrm{dBm}$ with the calculated result, and highlights these transitions. As in Fig.~\ref{fig:fig2}(b), a finite thermal occupation $n_{\rm th}=0.010$ is required to reproduce the data.}
\end{figure}

To probe the multiphoton transitions predicted by the JC model, we measured the readout-channel transmission $S_{21}$ as a function of drive detuning $\Delta=(\omega_d-\omega_c)/2\pi$ and drive power (Figs.~\ref{fig:fig2} and \ref{fig:fig3}). We first focus on the narrow detuning range shown in Fig.~\ref{fig:fig2}(b), where two series of peaks appear on the negative- and positive-detuning sides. The series are slightly asymmetric and, as anticipated from the dispersive effect of higher transmon levels, are collectively shifted toward positive detuning from the bare-cavity frequency. Our calculations identify these peaks as distinct multiphoton transitions, as labeled beside the corresponding spectral features. On the negative- (positive-) detuning side, transitions from the ground state $\lvert g,0\rangle$ to the dressed states $\lvert n,-\rangle$ ($\lvert n,+\rangle$) are resolved up to $n=20$ ($14$)~\cite{SuppMat}. The peaks broaden under strong driving because of saturation. This broadening is accompanied by the supersplitting observed in conventional transmission spectroscopy~\cite{bishopNonlinearResponseVacuum2009}.
\begin{figure*}[htbp]
    \includegraphics[width=0.93\textwidth]{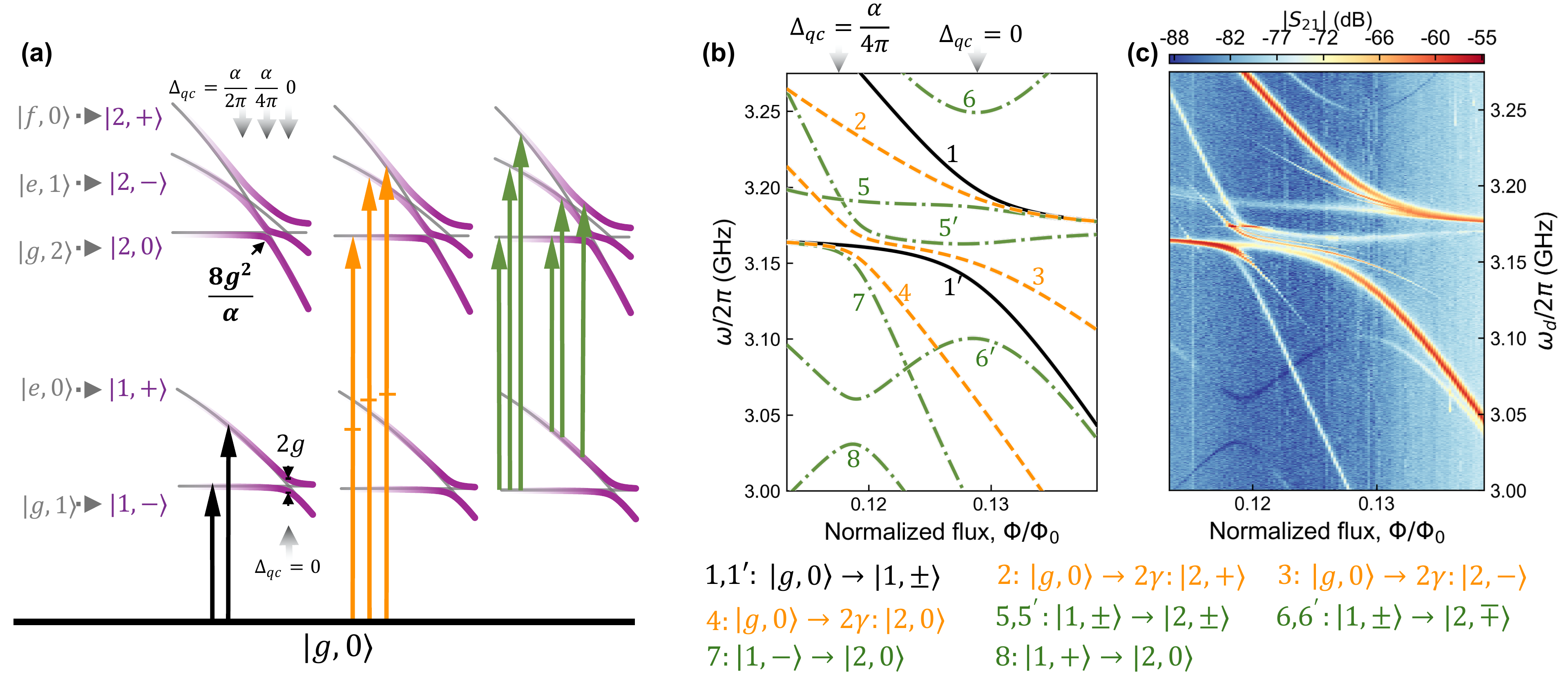}
    \caption{\label{fig:fig4}\textbf{QND-based spectroscopy for the system with qubit--cavity detuning.}
    \textbf{(a)} Energy-level diagram as a function of qubit flux, showing how the product states $\lvert g,1\rangle,\ldots,\lvert f,0\rangle$ evolve into the dressed states $\lvert 1,-\rangle,\ldots,\lvert 2,+\rangle$ through qubit--cavity coupling. Here, $\lvert f\rangle$ denotes the second excited state of the transmon, modeled as a Duffing oscillator. Black and yellow arrows indicate one- and two-photon transitions from the ground state, respectively. Green arrows indicate one-photon transitions originating from dressed states.
    \textbf{(b)} Calculated transition frequencies corresponding to the processes marked in panel (a), plotted as functions of normalized qubit flux. The assignments are shown at the bottom.
    \textbf{(c)} QND readout-channel transmission $S_{21}$ as a function of normalized qubit flux $\Phi/\Phi_0$ and drive frequency $\omega_d/2\pi$ at fixed drive power $P_d=-40~\mathrm{dBm}$. Its features agree perfectly with the theoretical prediction in (b).}
\end{figure*}
Transitions involving up to 6 photons were previously observed in Ref.~\cite{bishopNonlinearResponseVacuum2009}, whereas here we resolve transitions involving more than three times that number. For comparison, the inset in Fig.~\ref{fig:fig2}(b) shows the conventional cavity-transmission spectrum measured under the same conditions, in which the multiphoton transitions are obscured by the strong background of the drive field itself.
Besides the conventional transitions, we also observe multiphoton transitions from the lower dressed states $\lvert 1,-\rangle$ ($\lvert 1,+\rangle$) to higher dressed states $\lvert n,-\rangle$ ($\lvert n,+\rangle$), involving more than seven photons within the black box in Fig.~\ref{fig:fig2}(b). A finely sampled QND spectrum of this region is provided in the SM \cite{SuppMat}. These transitions eventually merge into two broad bands near zero detuning. We attribute the bands to overlapping transitions originating from successively higher dressed states, $\lvert 2,\pm\rangle$, $\lvert 3,\pm\rangle$, and so on. Numerical simulations show that their visibility depends critically on the finite thermal occupation of the cavity. Figure~\ref{fig:fig2}(d) compares experimental line cuts (dots) at drive powers of $-60$, $-50$, and $-40~\mathrm{dBm}$ with calculations for $n_{\rm th}=0$ (dashed curves) and $0.010$ (solid curves). Owing to a total attenuation of about $85~\mathrm{dB}$ in the experimental input line~\cite{SuppMat},
a bare cavity driven on resonance at these powers would contain
$\bar{n}=2P_{\rm in}/(\kappa\hbar\omega_{\rm r})\approx7.1$, $71$, and $710$ photons
at $P_d=-60$, $-50$, and $-40~\mathrm{dBm}$, respectively. The $n_{\rm th}=0$ calculation underestimates the background, and the peaks associated with $\lvert 1,\pm\rangle\to\lvert n,\pm\rangle$ have extremely low intensity. In contrast, the calculation with $n_{\rm th}=0.010$ agrees well with the data, and the corresponding peaks are enhanced by approximately $50$ and $25~\mathrm{dB}$ at $-60$ and $-50~\mathrm{dBm}$, respectively. These results show that a finite thermal population is essential for resolving these additional transitions and identifying the origin of the broadened bands. We note that the multiphoton transitions originating from $\lvert 1,\pm\rangle$ were not resolved in earlier studies.

To solidify the attribution of the peaks to multiphoton transitions, we manually selected the experimental transition frequencies at the centers of the resolved ridges in the two-dimensional spectrum. Because several features partially overlap or are asymmetric, we did not impose a common parametric line shape on all peaks. In Fig.~\ref{fig:fig2}(e), we plot the selected frequencies and their detunings from the cavity as functions of the photon number $n$ of the final dressed states, and compare them with predictions of the full multilevel model. The good agreement between experiment and theory further supports our assignment of the spectral features to the corresponding multiphoton transitions.

Figure~\ref{fig:fig3} extends the QND-based spectrum over a wider drive-detuning range and reveals two additional weak features at large negative and positive detunings. Comparing their positions with the transitions shown on the right side of Fig.~\ref{fig:fig2}(a), we assign the negative- (positive-) detuning feature to the transition $\lvert 1,+\rangle\to\lvert 2,-\rangle$ ($\lvert 1,-\rangle\to\lvert 2,+\rangle$) between states of opposite parity. Although these transitions are suppressed in the idealized parity-selection picture, perturbation theory predicts a small but nonzero transition matrix element. This accounts for their weak intensity and the larger drive power required relative to same-parity transitions such as $\lvert 1,\pm\rangle\to\lvert n,\pm\rangle$. To our knowledge, these transitions have not been reported previously. They are resolved here through the combination of reduced background in QND-based spectroscopy and finite thermal population of the dressed states.

Having observed multiphoton transitions originating from dressed states, including both same- and opposite-parity transitions, we now examine their dependence on the qubit--cavity detuning $\Delta_{qc}$. We generalize the energy-level diagram in Fig.~\ref{fig:fig2}(a) to finite $\Delta_{qc}$ by tuning the qubit frequency with magnetic flux, as shown in Fig.~\ref{fig:fig4}(a). When $\Delta_{qc}$ becomes comparable to the qubit anharmonicity $\alpha/2\pi$, multiphoton processes involving the second excited state $\lvert f\rangle$ of the transmon must be included. In the left part of Fig.~\ref{fig:fig4}(a), as $\Delta_{qc}$ decreases from left to right, the nearly uncoupled product states at large detuning (gray) evolve into dressed states at small detuning (purple). The uncoupled two-excitation subspace has degenerate pairs $\lvert e,1\rangle$--$\lvert f,0\rangle$, $\lvert g,2\rangle$--$\lvert f,0\rangle$, and $\lvert g,2\rangle$--$\lvert e,1\rangle$ at $\Delta_{qc}=\alpha/2\pi$, $\alpha/4\pi$, and $0$, respectively. The first and third pairs are directly connected by single-photon exchange with coupling strength $\sqrt{2}g$, giving an avoided-crossing splitting of approximately $2\sqrt{2}g$. At $\Delta_{qc}=\alpha/4\pi$, by contrast, $\lvert g,2\rangle$ and $\lvert f,0\rangle$ have no direct coupling matrix element. However, the off-resonant state $\lvert e,1\rangle$ mediates a virtual two-photon exchange and opens a splitting of $\sqrt{(\alpha/4)^2+4g^2}-\alpha/4$.

The enlarged energy-level diagram reveals a rich set of transitions [Figs.~\ref{fig:fig4}(a) and \ref{fig:fig4}(b)]. One-photon transitions from the ground state $\lvert g,0\rangle$ to $\lvert 1,-\rangle$ and $\lvert 1,+\rangle$ form an avoided crossing at $\Delta_{qc}=0$ (black arrows and curves). Two-photon transitions from $\lvert g,0\rangle$ to $\lvert 2,r\rangle$ (yellow arrows and curves) and one-photon transitions from the excited dressed states $\lvert 1,s\rangle$ to $\lvert 2,r\rangle$ (green arrows and curves), with $s=\pm$ and $r=-,0,+$, probe the same two-excitation eigenenergies. They therefore inherit avoided crossings at $\Delta_{qc}=\alpha/4\pi$ and $0$ [the crossing at $\alpha/2\pi$ is outside the range shown in Fig.~\ref{fig:fig4}(b)]. Because the two sets have different initial states and photon orders, their avoided crossings occur at different transition frequencies and exhibit different spectral splittings. Near $\Delta_{qc}=0$, the spectrum also contains two nearly parallel, arch-like transitions $\lvert 1,\pm\rangle\to\lvert 2,\pm\rangle$ between states of the same parity and two oppositely oriented, parabolic transitions $\lvert 1,\pm\rangle\to\lvert 2,\mp\rangle$ between states of different parity. The former were reported previously \cite{finkClimbingJaynesCummings2008a,bonsenProbingJaynesCummingsLadder2023}, whereas the latter have not been reported before.

Figure~\ref{fig:fig4}(c) shows excellent agreement between the measured spectrum and the calculated transition frequencies in Fig.~\ref{fig:fig4}(b). The readout-channel transmission $S_{21}$ was recorded as a function of qubit flux and drive frequency at a fixed drive power of $-40~\mathrm{dBm}$. The one-photon transitions from the ground state are strongest, whereas the two-photon transitions have intermediate intensity because of their smaller transition matrix elements. One-photon transitions originating from lower dressed states are weaker because those states have smaller populations. Among these transitions, those between states of the same parity are stronger than those between states of different parity because their matrix elements are larger. The transition intensities also evolve asymmetrically with detuning, reflecting changes in the qubit component of the dressed states.

In summary, we have developed QND-based spectroscopy with a separate cavity drive and qubit readout. This approach resolves multiphoton transitions in a strongly driven circuit-QED system that are obscured in conventional transmission measurements. Finite thermal populations provide access to transitions originating from excited JC dressed states, allowing us to identify the near-resonant features as transitions between states of the same parity, and to resolve weaker transitions between states of different parity. Tuning the qubit--cavity detuning further reveals the role of the transmon multilevel structure, including avoided crossings involving its second excited state and a virtual-photon-mediated interaction. These results establish QND-based spectroscopy as a sensitive probe of the JC ladder. In the future, with sufficient measurement bandwidth, a time-resolved implementation could reveal individual quantum jumps between dressed states and determine state-to-state transition and relaxation rates. Such measurements could connect microscopic dressed-state dynamics to bistability and the quantum-to-classical crossover in driven light--matter systems.

\begin{acknowledgments}
We thank Tao Jiang and Naibin Zhou for their help in developing the hardware and software during the early stage of the experiment, and Profs. William John Munro and Yizheng Zhen for fruitful discussions. We are grateful to Profs. Jian-Wei Pan, and Xiaobo Zhu for their crucial support of this project. This work was supported by the Quantum Science and Technology--National Science and Technology Major Project (No. 2024ZD0301200), the National Natural Science Foundation of China (Nos. W2431001, 12422413, and 62475242), the Scientific Research Innovation Capability Support Project for Young Faculty (No. SRICSPYF-BS2025008), and the Natural Science Foundation of Henan (No. 252300421301).
\end{acknowledgments}

\bibliography{ref}

%%%%%%%%%%%%%%%% Supplementary Material
\clearpage
% ---- 补充材料计数器：重置并加 S 前缀 ----
\setcounter{section}{0}
\renewcommand{\thesection}{S\arabic{section}}
\renewcommand{\thesubsection}{\thesection.\arabic{subsection}}
\setcounter{figure}{0}
\renewcommand{\thefigure}{S\arabic{figure}}
\setcounter{table}{0}
\renewcommand{\thetable}{S\arabic{table}}
\setcounter{equation}{0}
\renewcommand{\theequation}{S\arabic{equation}}
\renewcommand{\theHsection}{S\arabic{section}}
\renewcommand{\theHsubsection}{\theHsection.\arabic{subsection}}
\renewcommand{\theHfigure}{S\arabic{figure}}
\renewcommand{\theHtable}{S\arabic{table}}
\renewcommand{\theHequation}{S\arabic{equation}}

\begin{center}
{\Large\bfseries Supplemental Material}\\[2mm]
\end{center}

\lstdefinestyle{qutipcode}{
    language=Python,
    basicstyle=\ttfamily\fontsize{6.4}{7.4}\selectfont,
    keywordstyle=\color{black},
    commentstyle=\color{black},
    stringstyle=\color{black},
    numbers=left,
    numberstyle=\ttfamily\fontsize{6.4}{7.4}\selectfont,
    numbersep=7pt,
    stepnumber=1,
    backgroundcolor=\color{codebackground},
    frame=single,
    rulecolor=\color{codeframe},
    framerule=0.5pt,
    framesep=4pt,
    xleftmargin=2.2em,
    framexleftmargin=1.9em,
    framexrightmargin=0.4em,
    framextopmargin=3pt,
    framexbottommargin=3pt,
    breaklines=true,
    breakatwhitespace=true,
    columns=fullflexible,
    keepspaces=true,
    showstringspaces=false,
    tabsize=4,
    aboveskip=0pt,
    belowskip=0pt
}
\renewcommand{\thesection}{S\arabic{section}}
\renewcommand{\thesubsection}{\thesection.\arabic{subsection}}
\renewcommand{\thefigure}{S\arabic{figure}}
\renewcommand{\thetable}{S\arabic{table}}
\renewcommand{\theequation}{S\arabic{equation}}
\setcounter{secnumdepth}{2}

\newcommand{\nth}{n_{\rm th}}
\newcommand{\Diss}[1]{\mathcal{D}\!\left[#1\right]}
% Outline
\setcounter{tocdepth}{2}
\tableofcontents

The main text presents the central physics and experimental observations. This Supplemental Material provides further details of the experimental implementation, data analysis, 
theoretical model and numerical methods.  Section~\ref{sec:device} describes the experimental platform, readout calibration, spectroscopy comparison, high-resolution spectra, and data processing. Section~\ref{sec:model} derives the multilevel eigenenergies and transition frequencies using fixed-excitation blocks of the Hamiltonian. Section~\ref{sec:numerics} describes the QuTiP steady-state simulations and their comparison with experiments. Section~\ref{sec:flux_transitions} analyzes the flux-dependent transitions and the virtual-photon-mediated two-photon exchange in Fig.~4.

\section{Experimental platform and methods}
\label{sec:device}

This section provides the experimental details underlying Fig.~1 and the measurements in Figs.~2--4 of the main text. We first describe the on-chip implementation and chip parameters, and then describe the external microwave connections and measurement protocol. Finally, we compare the QND readout with direct cavity spectroscopy and document the processing used for the Fig.~2 frequency analysis in the main text.

\begin{figure}[bp]
\centering
\includegraphics[width=\columnwidth]{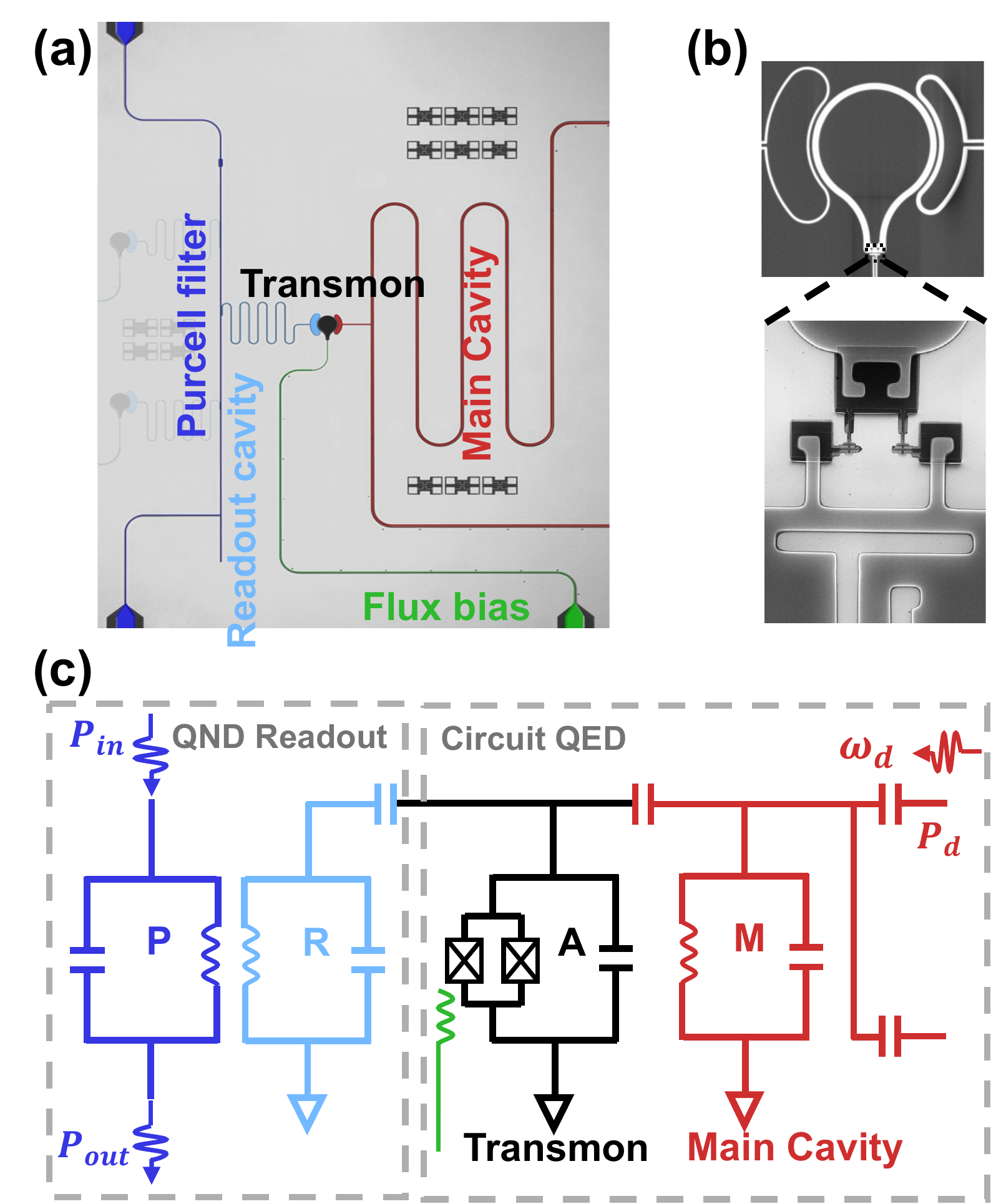}
\caption{Details on the circuit-QED and QND readout system.
    \textbf{(a)} False-color optical micrograph of the fabricated chip. The colors were added to distinguish the functional circuit elements. \textbf{(b)} Scanning electron micrographs of the transmon junction region. The upper image shows the SQUID together with the transmon-capacitor structure, and the lower image provides a magnified view of the SQUID loop and Josephson-junction region. \textbf{(c)} Equivalent circuit of the fabricated device. The left block contains the readout resonator and the inline Purcell-filter channel whose transmission $S_{21}$ is measured for QND readout, whereas the right block contains the driven main-cavity--transmon circuit-QED subsystem and its flux-bias line.}
\label{fig:chip_circuit}
\end{figure}

\subsection{Chip circuit and fabrication}

Figure~\ref{fig:chip_circuit} shows the physical implementation and equivalent
circuit of the separated drive--readout scheme introduced in Fig.~1 of the
main text. The false-color optical micrograph in Fig.~\ref{fig:chip_circuit}(a)
  identifies the main cavity, transmon, readout resonator, Purcell filter~\cite{jeffreyFastAccurateState2014}, and
flux-bias line on the fabricated chip. Figure~\ref{fig:chip_circuit}(b) shows
the SQUID and Josephson-junction region forming the flux-tunable transmon.
The equivalent circuit in Fig.~\ref{fig:chip_circuit}(c) makes the two
functional subsystems explicit: the main-cavity--transmon circuit used for the spectroscopy drive and the readout-resonator--Purcell-filter channel used for QND readout. The simplified schematic in Fig.~1 of the main text omits the inline Purcell filter. In the physical circuit, a microwave field probes the Purcell-filter feedline, and the transmission $S_{21}$ is measured for the transmon-state QND readout.

The device was fabricated on a $430~\mu\mathrm{m}$-thick sapphire substrate. A $200~\mathrm{nm}$ tantalum film was patterned by laser direct-write lithography and reactive-ion etching to define the coplanar-waveguide transmission lines, the main cavity, and the readout resonator. The Al/AlO$_x$/Al Josephson junctions were fabricated by Dolan-bridge double-angle evaporation with an intermediate oxidation step~\cite{dolanOffsetMasksLiftoff1977}. The chip was mounted in a sample holder and wire bonded, with additional ground bonds placed across the coplanar-waveguide gaps to suppress slotline modes.

Table~\ref{tab:hamiltonian_parameters} lists the parameters of both the driven circuit-QED subsystem and the independent QND readout subsystem. The effective dissipation parameters used in the numerical simulations are listed separately in Table~\ref{tab:steady_state_parameters}.

\subsection{Cryogenic setup and measurement protocol}

The chip was mounted and thermalized at the base stage of a dilution refrigerator at approximately $13~\mathrm{mK}$. A complete diagram of the cryogenic microwave wiring and room-temperature instrumentation, including the pulsed and CW readout configurations, is provided in Fig.~\ref{fig:setup} at the end of this Supplemental Material. The spectroscopy drive tone and readout probe are delivered through separate microwave lines. At room temperature, the readout channel supports two measurement configurations. A pulsed DAC/ADC chain is used together with the dedicated transmon-control line for standard qubit calibrations and transmon-state preparation with calibrated control pulses, whereas a vector network analyzer (VNA) is used for continuous-wave (CW) steady-state transmission spectroscopy.

\begin{figure}[htbp]
\centering
\includegraphics[width=0.8\linewidth]{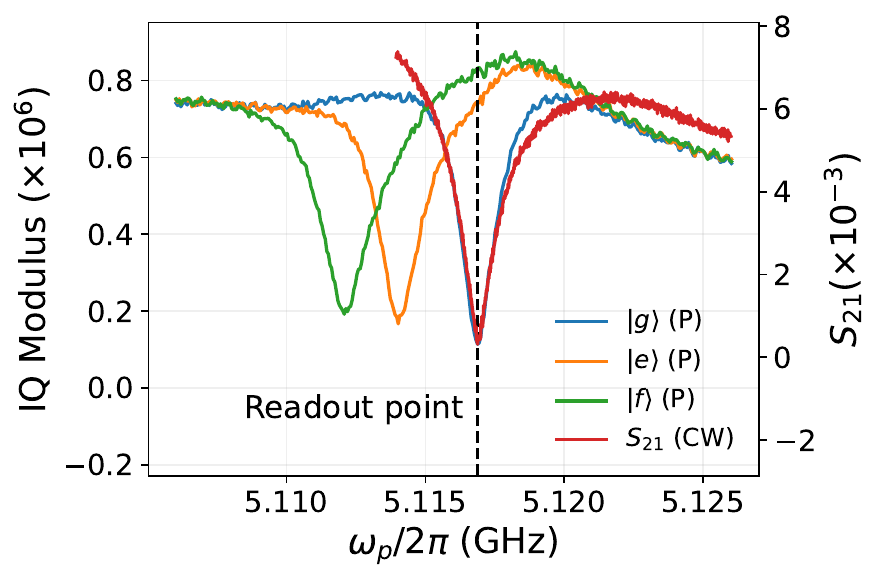}
\caption{Readout-resonator response as a function of probe frequency $\omega_p/2\pi$. The curve labeled $S_{21}$ (CW; right axis) shows the continuous-wave transmission amplitude. The blue, orange, and green curves (P; left axis) show the pulsed-readout IQ modulus for the transmon prepared in $|g\rangle$, $|e\rangle$, and $|f\rangle$, respectively. The vertical gray dashed line marks the fixed CW readout point used for the spectroscopy measurements. At this frequency, the measured transmission is sensitive to the transmon excitation and is used as an experimental proxy for $\langle\hat b^\dagger\hat b\rangle$.}
\label{fig:readout_response}
\end{figure}

Figure~\ref{fig:readout_response} compares the readout-resonator responses obtained with the pulsed and CW configurations. In the pulsed-readout (P) measurement, the transmon is prepared in $|g\rangle$, $|e\rangle$, and $|f\rangle$, and the corresponding state-dependent responses are recorded. In the CW measurement, the transmon remains predominantly in $|g\rangle$, so the CW response is consistent with the $|g\rangle$ pulsed-readout trace. We then choose the CW probe frequency at the point marked in Fig.~\ref{fig:readout_response}. At this operating point, the separation of the state-resolved pulsed responses shows that the transmission amplitude is sensitive to the transmon excitation. During the spectroscopy measurements, variations in the CW transmission amplitude are therefore used as an experimental proxy for the mean steady-state transmon occupation $\langle\hat b^\dagger\hat b\rangle$. All spectroscopy measurements reported in the main text and below use the CW configuration.

Because the readout resonator is dispersively coupled to the flux-tunable transmon, its resonance frequency changes when the dc flux-bias voltage $V_{\rm dc}$ changes the transmon frequency. The readout resonator is not directly flux biased, but its resonance is shifted through this coupling. The readout probe frequency must therefore be calibrated as $V_{\rm dc}$ is varied. We first summarize the dispersive model and then describe the experimental calibration.

\begin{table*}[htbp]
\caption{Parameters of the circuit-QED spectroscopy and QND readout subsystems.  The readout-resonator--transmon coupling is inferred from the measured dispersive shift, while the Purcell-filter parameters are design values.}
\label{tab:hamiltonian_parameters}
\centering
\begin{ruledtabular}
\begin{tabular}{llcc}
Subsystem & Quantity & Symbol & Value \\
Circuit QED & Main-cavity frequency & $\omega_c/2\pi$ & $3.171~\mathrm{GHz}$ \\
& Main-cavity linewidth (measured)  & $\kappa/2\pi$ & $67~\mathrm{kHz}$ \\
& Transmon tuning range & $\omega_q/2\pi$ & $2.800\text{--}3.856~\mathrm{GHz}$ \\
& Cavity--transmon coupling & $g/2\pi$ & $31~\mathrm{MHz}$ \\
& Transmon anharmonicity magnitude & $\alpha/2\pi$ & $192~\mathrm{MHz}$ \\
& Transmon relaxation time (measured) & $T_1$ & $38~\mu\mathrm{s}$ \\
& Ramsey coherence time (measured) & $T_2^*$ & $4.6~\mu\mathrm{s}$ \\
& Echo coherence time (measured) & $T_{2,\mathrm{echo}}$ & $5.2~\mu\mathrm{s}$ \\
QND readout & Readout-resonator frequency (measured) & $\omega_{\rm r}/2\pi$ & $5.117~\mathrm{GHz}$ \\
& Readout-resonator linewidth (measured) & $\kappa_{\rm r}/2\pi$ & $2.3~\mathrm{MHz}$ \\
& Transmon frequency during characterization & $\omega_q/2\pi$ & $3.856~\mathrm{GHz}$ \\
& Dispersive-shift magnitude (measured) & $|\chi|/2\pi=(\chi_0-\chi_1)/2\pi$ & $1.5~\mathrm{MHz}$ \\
& Readout-resonator--transmon coupling (inferred) & $g_{\rm r}/2\pi$ & $120~\mathrm{MHz}$ \\
& Purcell-filter center frequency (design) & $\omega_{\rm P}/2\pi$ & $\sim5.1~\mathrm{GHz}$ \\
& Purcell-filter linewidth (design) & $\kappa_{\rm P}/2\pi$ & $\sim200~\mathrm{MHz}$ \\
\end{tabular}
\end{ruledtabular}
\end{table*}

We define the transmon--readout-resonator detuning as $\Delta_{\rm qr}(\Phi)=\omega_q(\Phi)-\omega_{\rm r}^{(0)}$, where $\omega_q(\Phi)$ is the flux-dependent transmon frequency and $\omega_{\rm r}^{(0)}$ is the uncoupled readout-resonator frequency. The dispersive approximation applies when $|\Delta_{\rm qr}|\gg g_{\rm r}$ and $|\Delta_{\rm qr}-\alpha|\gg\sqrt{2}\,g_{\rm r}$. Here $g_{\rm r}$ is the transmon--readout-resonator coupling. The anharmonicity parameter satisfies $\alpha>0$ and $\omega_{ef}=\omega_q-\alpha$, while $\Phi$ denotes the flux through the transmon SQUID. 

At the characterization point, Table~\ref{tab:hamiltonian_parameters} gives $\omega_q/2\pi=3.856~\mathrm{GHz}$, $\omega_{\rm r}^{(0)}/2\pi=5.117~\mathrm{GHz}$, $g_{\rm r}/2\pi=120~\mathrm{MHz}$, and $\alpha/2\pi=192~\mathrm{MHz}$. These values give $\Delta_{\rm qr}/2\pi=-1.261~\mathrm{GHz}$ and therefore $|\Delta_{\rm qr}|/2\pi\simeq 10.5\,(g_{\rm r}/2\pi)$. Including the anharmonicity, $|\Delta_{\rm qr}-\alpha|/2\pi=1.453~\mathrm{GHz}\simeq8.6\,[(\sqrt{2}\,g_{\rm r})/2\pi]$. Both inequalities are well satisfied, confirming that the readout measurement operates in the dispersive regime. In this regime, real excitation exchange is suppressed and the readout photon only produces a virtual, transmon-state-dependent frequency shift~\cite{blaisCavityQuantumElectrodynamics2004}. The readout subsystem is therefore described by the dispersive Hamiltonian:
\begin{equation}
\frac{\hat H_{\rm disp}}{\hbar}
=
\omega_{\rm r}^{(0)}\hat a_{\rm r}^{\dagger}\hat a_{\rm r}
+\sum_j\left[\widetilde{\omega}_j(\Phi)+\chi_j(\Phi)\hat a_{\rm r}^{\dagger}\hat a_{\rm r}\right]|j\rangle\langle j|,
\label{eq:readout_subsystem}
\end{equation}
The first term gives the bare readout-resonator energy in angular-frequency units. The operator $\hat a_{\rm r}$ annihilates a readout-resonator photon, so $\hat a_{\rm r}^{\dagger}\hat a_{\rm r}$ is the photon-number operator. The sum runs over the transmon levels $j=0,1,\ldots$, and $|j\rangle\langle j|$ projects onto level $j$. The quantity $\widetilde{\omega}_j(\Phi)$ is the corresponding flux-dependent level energy divided by $\hbar$. The term $\chi_j(\Phi)\hat a_{\rm r}^{\dagger}\hat a_{\rm r}$ gives the dispersive shift per readout photon for that level.

To leading order, the state-dependent dispersive shift and the corresponding readout-resonator resonance frequency are \cite{blaisCavityQuantumElectrodynamics2004}
\begin{equation}
\begin{aligned}
\chi_j(\Phi)
&=g_{\rm r}^{2}\left[
\frac{j}{\Delta_{\rm qr}(\Phi)-\alpha(j-1)}
-\frac{j+1}{\Delta_{\rm qr}(\Phi)-\alpha j}
\right],
\\
\omega_{{\rm r},j}(\Phi)
&=\omega_{\rm r}^{(0)}+\chi_j(\Phi).
\end{aligned}
\label{eq:readout_dispersive_shift}
\end{equation}
Thus, the readout-resonator resonance frequency depends on the transmon state and is given by $\omega_{{\rm r},j}$.

\begin{figure}[htbp]
\centering
\includegraphics[width=0.8\linewidth]{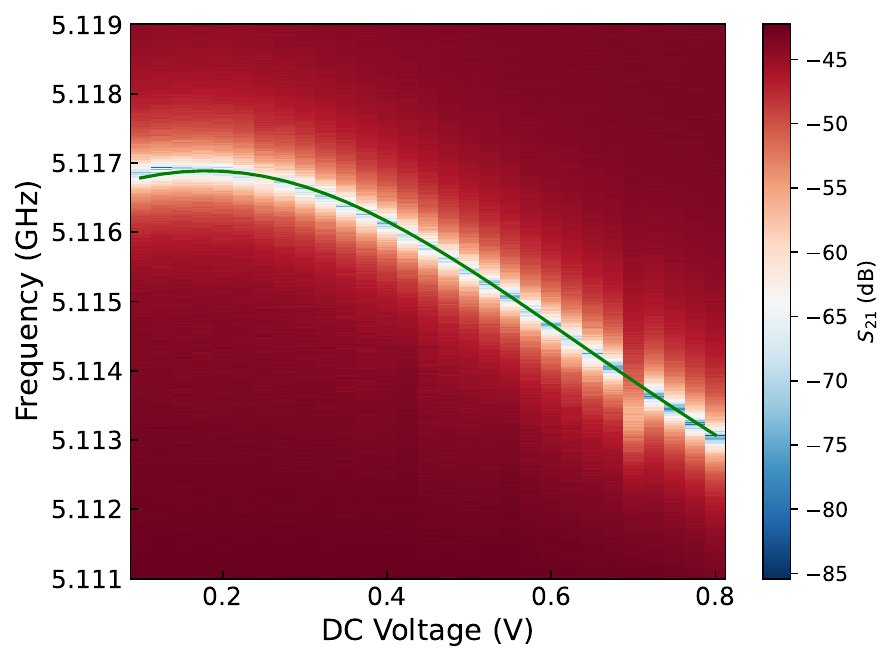}
\caption{Readout-resonator transmission $S_{21}$ as a function of probe frequency $\omega_p/2\pi$ and dc flux-bias voltage $V_{\rm dc}$. At each bias, the readout-resonator resonance frequency is obtained from a Lorentzian fit to the corresponding frequency trace. The green curve is a fit to the extracted resonance frequencies and defines the flux-dependent CW readout point used in the spectroscopy measurements.}
\label{fig:S21ZDC}
\end{figure}

For the calibration (Fig.~\ref{fig:S21ZDC}), the readout tone is applied continuously and the transmitted field is measured after the system reaches steady state. Over the narrow frequency range of one trace, the complex transmission of the hanger-type readout channel is
\begin{equation}
S_{21}^{(j)}(\omega_p,\Phi)
\simeq
S_{\rm bg}(\omega_p)
+
\frac{A_{\rm r}(\omega_p)}
{\dfrac{\kappa_{\rm r}}{2}
-i\left[\omega_p-\omega_{{\rm r},j}(\Phi)\right]},
\label{eq:readout_s21_hanger}
\end{equation}
Here $S_{\rm bg}$ is the slowly varying background, $A_{\rm r}$ contains the coupling and measurement-chain factors, and $\kappa_{\rm r}$ is the readout-resonator linewidth. The resonant term is centered at $\omega_{{\rm r},j}(\Phi)$. During the weak CW calibration, the transmon remains predominantly in $|g\rangle$. The measured resonance therefore follows $\omega_{\rm r}(V_{\rm dc})\simeq\omega_{\rm r}^{(0)}+\chi_0[\Phi(V_{\rm dc})]$. Since $\chi_0=-g_{\rm r}^2/\Delta_{\rm qr}$, this relation is equivalently $\omega_{\rm r}(V_{\rm dc})\simeq\omega_{\rm r}^{(0)}-g_{\rm r}^2/[\omega_q(\Phi(V_{\rm dc}))-\omega_{\rm r}^{(0)}]$.

The dc bias changes the SQUID flux and therefore the transmon frequency. For a nearly symmetric SQUID, this dependence is $\omega_q(\Phi)\simeq[\sqrt{8E_CE_{J,\max}|\cos(\pi\Phi/\Phi_0)|}-E_C]/\hbar$. Here $E_C$ is the charging energy, $E_{J,\max}$ is the maximum Josephson energy, and $\Phi_0$ is the flux quantum. The bias therefore changes the readout-resonator resonance through the chain $V_{\rm dc}\!\rightarrow\Phi\!\rightarrow\omega_q\!\rightarrow\Delta_{\rm qr}\!\rightarrow\chi_0\!\rightarrow\omega_{\rm r}$. We determine this dependence experimentally with weak CW VNA scans. At each $V_{\rm dc}$, we acquire a frequency trace at a readout power of $-30~\mathrm{dBm}$. We extract $f_{\rm r}(V_{\rm dc})=\omega_{\rm r}(V_{\rm dc})/2\pi$ from a Lorentzian fit to each trace. Figure~\ref{fig:S21ZDC} shows the resulting map. The smooth motion of this curve with $V_{\rm dc}$ is the measured flux-mediated resonance shift described above.

Because the transmon remains below the readout frequency in this calibration range, $\Delta_{\rm qr}<0$ and the ground-state dispersive shift is positive. Combining the SQUID dependence with the relation above gives an inverse dependence of the readout-resonator resonance frequency on $V_{\rm dc}$. We therefore parameterize the extracted resonance frequencies by
\begin{equation*}
f_{\rm r}^{\rm fit}(V_{\rm dc})
=
f_{\rm bare}
+
\frac{1}
{b-c\sqrt{\left|\cos\!\left[a(V_{\rm dc}-V_0)\right]\right|}},
\end{equation*}
where $a$, $b$, $c$, $V_0$, and $f_{\rm bare}$ are fit parameters. The parameters $a$ and $V_0$ absorb the flux-to-voltage conversion and the bias offset, while the remaining parameters absorb frequency scales and offsets. The green curve is the fitted curve, which follows the extracted resonance frequencies. During the subsequent spectroscopy scans, the main-cavity drive is swept independently while the readout power is held fixed, using the calibrated probe frequency at each $V_{\rm dc}$.

\begin{figure}[htbp]
\centering
\includegraphics[width=0.8\linewidth]{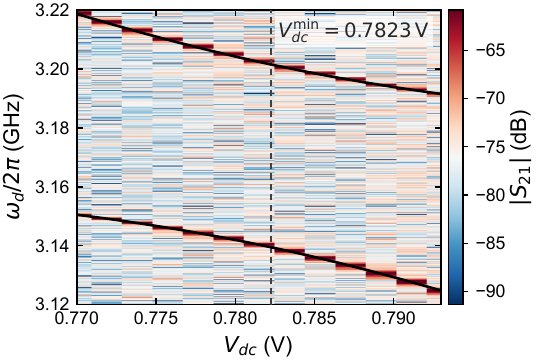}
\caption{Low-power qubit--cavity avoided-crossing spectrum. The color scale shows the measured readout-resonator transmission $S_{21}$ as a function of spectroscopy-drive frequency $\omega_d/2\pi$ and dc flux-bias voltage $V_{\rm dc}$. The black curves fit the upper and lower dressed-mode frequencies. The vertical dashed line marks the bias at which their frequency separation is minimal; this bias defines the nominal qubit--cavity resonance point used for the measurements in Figs.~2 and 3 of the main text.}
\label{fig:AC_fit}
\end{figure}

The nominal qubit--cavity resonance is determined from the low-power avoided-crossing spectrum in Fig.~\ref{fig:AC_fit}, where multiphoton transitions are suppressed. The black curves fit the upper and lower dressed-mode frequencies. The vertical dashed line marks the flux-bias voltage at which their frequency separation is minimal; we take this bias as the experimental qubit--cavity resonance point, $\omega_q\simeq\omega_c$. This bias is used for the measurements in Figs.~2 and 3 of the main text. Because the resonance setting has finite experimental accuracy, a small residual detuning $\Delta_{qc}=(\omega_q-\omega_c)/2\pi$ is retained in the quantitative transition-frequency comparison, as described in the main text.

\subsection{Comparison of QND-based and direct cavity spectroscopy}
\label{sec:measurement_comparison}

\begin{figure*}[t]
\centering
\includegraphics[width=0.85\linewidth]{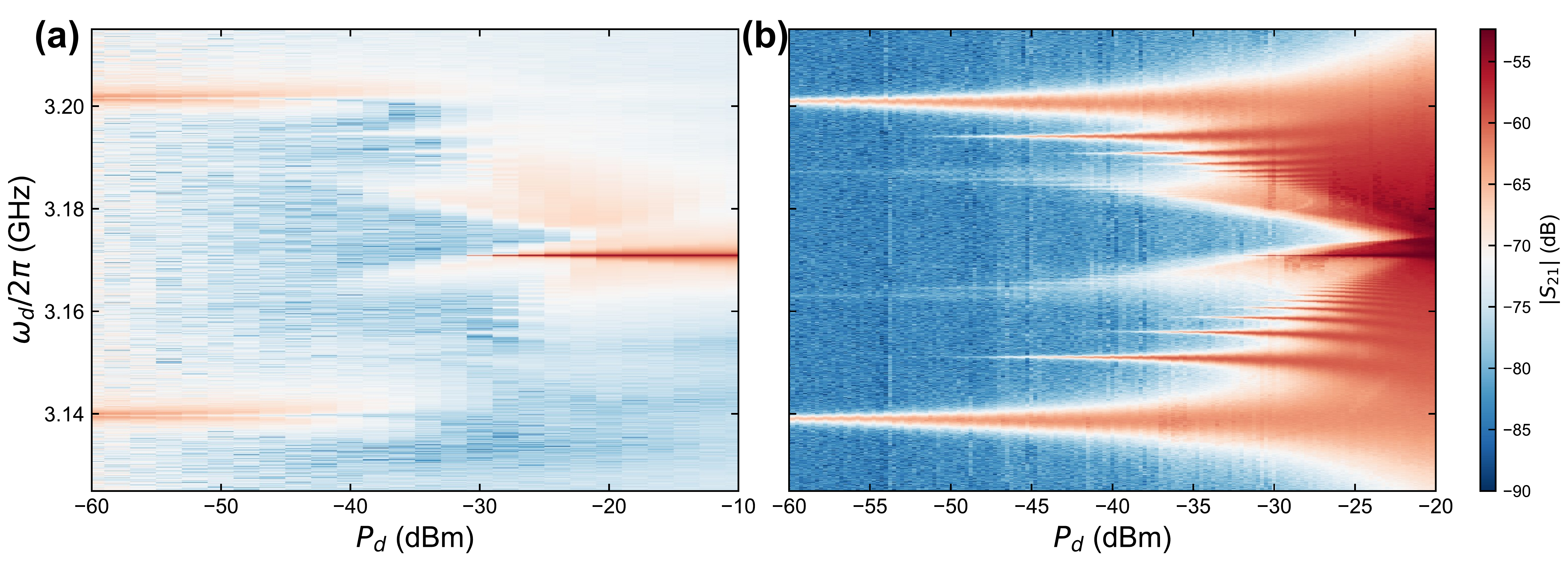}
\caption{Comparison of conventional transmission spectroscopy with QND-based spectroscopy over the same spectroscopy-drive frequency window and with the same frequency resolution. \textbf{(a)} Direct cavity-transmission spectroscopy. The drive power is sampled in $2~\mathrm{dB}$ steps, and the number of averages is increased from $2$ at high power $P_d = -10~\mathrm{dBm}$ to $500$ at $P_d = -60~\mathrm{dBm}$ to compensate for the decreasing transmitted signal. Even with this power-dependent averaging, the discrete thermally assisted transitions are difficult to resolve and are not discernible in the weak-drive regime. \textbf{(b)} QND-based spectroscopy acquired with $30$ averages at every drive power. The thermally assisted transitions remain well resolved with high contrast, even at low drive powers, because the readout strength is independent of $P_d$. Panel (a) is reproduced as the inset of Fig.~2(b) in the main text.}
\label{fig:measurement_comparison}
\end{figure*}

To quantify the advantage of our QND-based spectroscopy, Fig.~\ref{fig:measurement_comparison} compares power-dependent spectra measured by direct cavity transmission and by the separated QND readout. The two measurements use the same spectroscopy-drive frequency window and frequency resolution.

In direct cavity spectroscopy [Fig.~\ref{fig:measurement_comparison}(a)], the same field provides both the excitation and the measured transmission signal. Lowering the microwave power $P_d$ therefore also reduces the detected coherent field, requiring progressively more averaging to maintain a usable signal-to-noise ratio. The drive power on the left panel is sampled in steps of $2~\mathrm{dB}$, while the number of averages is increased from $2$ to $500$ as $P_d$ is reduced from $-10$ to $-60~\mathrm{dBm}$. This rapidly increasing averaging requirement leads to a very long acquisition time and motivates the relatively coarse power spacing. By contrast, the QND readout-based spectrum [Fig.~\ref{fig:measurement_comparison}(b)] is acquired with $30$ averages at every drive power because the readout strength is set independently of $P_d$.

Despite the much larger number of averages used for direct transmission at low powers, the left panel does not clearly resolve the thermally assisted response into discrete transitions, and the thermally assisted features are no longer discernible in the weak-drive regime. The QND readout result, in contrast, resolves the discrete thermally assisted transitions with substantially higher contrast and retains their visibility even at low drive powers. The comparison therefore shows that the improved weak-transition sensitivity arises from separating the readout strength from the spectroscopy-drive power, rather than from increasing the acquisition time.

\begin{figure*}[t]
\centering
\includegraphics[width=0.9\linewidth]{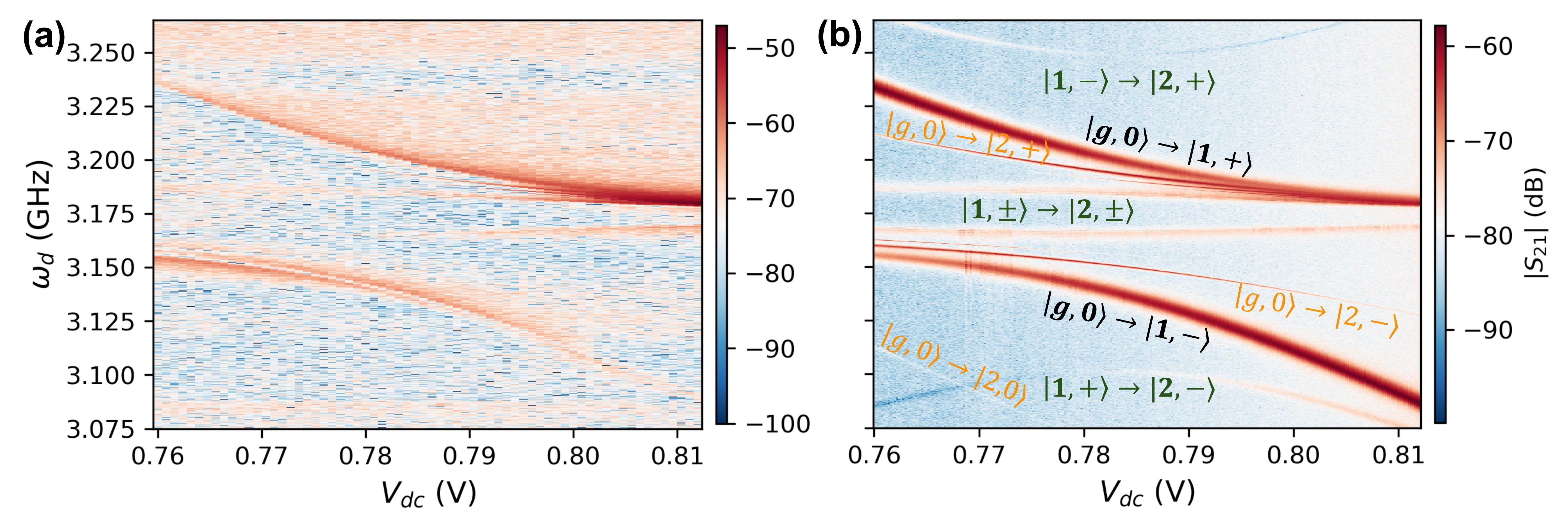}
\caption{Comparison of conventional transmission spectroscopy \textbf{(a)} and QND-based spectroscopy \textbf{(b)} at a fixed drive power $P_d=-40~\mathrm{dBm}$, with each data point averaged 50 times. The number of sampled points along the horizontal axis is one quarter of that in panel (b), whereas the number along the vertical axis is the same. Supersplitting prevents the transitions from being resolved into distinct lines, and opposite-parity transitions are not visible. In panel (b), the individual transition lines, including the weaker transitions between the opposite-parity states $|1,\pm\rangle\to|2,\mp\rangle$, remain visible.}
\label{fig:measurement_AC_comparison}
\end{figure*}

A direct comparison at a fixed drive power is shown in Fig.~\ref{fig:measurement_AC_comparison}. Both spectra were acquired at $P_d=-40~\mathrm{dBm}$, with each data point averaged 50 times. The direct cavity-transmission spectrum in Fig.~\ref{fig:measurement_AC_comparison}(a) contains one quarter as many points along the voltage axis as the QND spectrum in Fig.~\ref{fig:measurement_AC_comparison}(b), while the two measurements contain the same number of points along the frequency axis. In the direct measurement, supersplitting~\cite{bishopNonlinearResponseVacuum2009} causes the transition features to overlap, preventing the individual lines from being resolved, and the weaker opposite-parity transitions are not discernible. In contrast, the separated QND readout clearly resolves the individual transition lines, including the opposite-parity transitions. This fixed-power comparison further demonstrates the enhanced spectral resolution provided by separating the drive and readout channels.

\begin{figure}[htbp]
\centering
\includegraphics[width=\columnwidth]{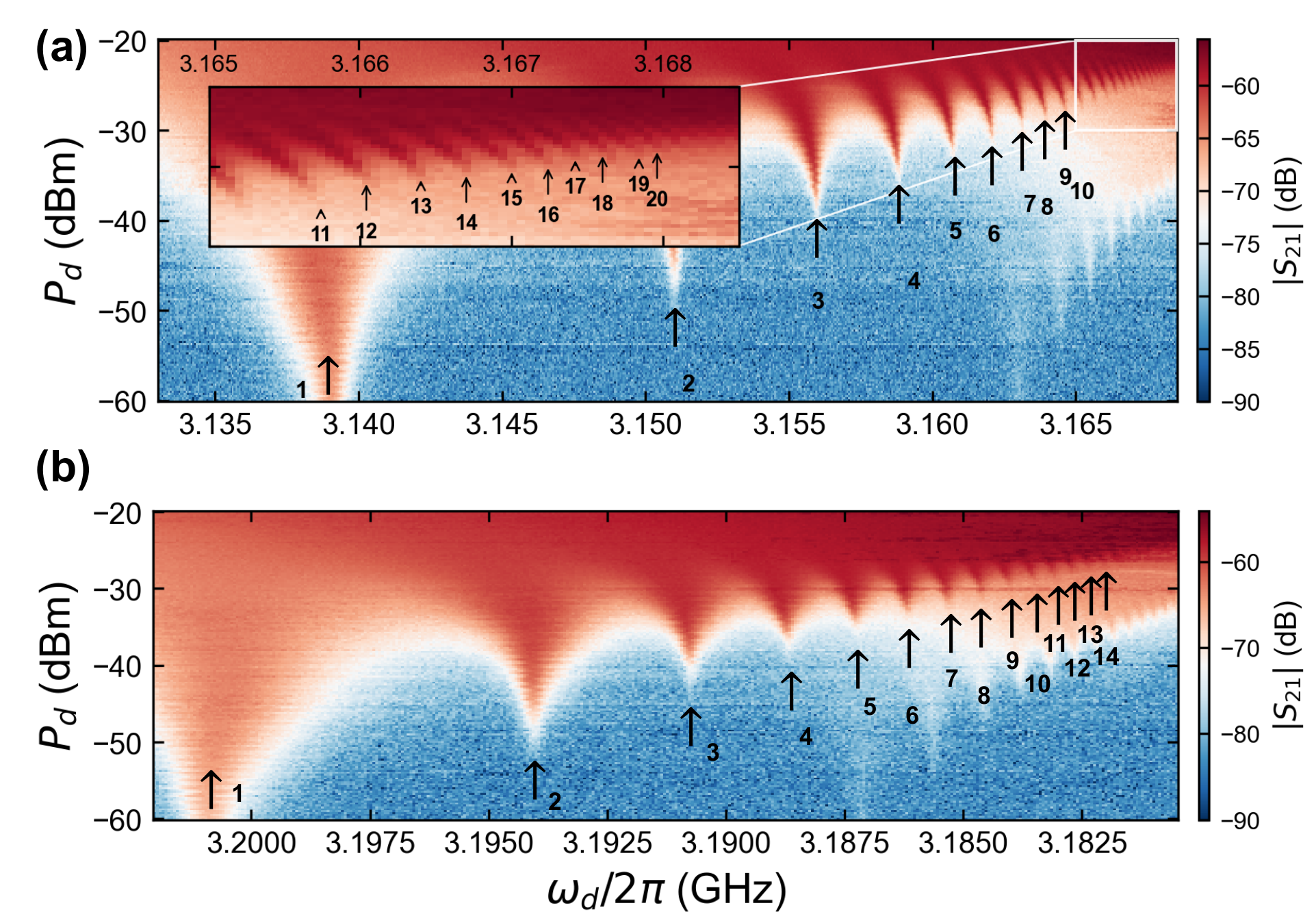}
\caption{Enlarged views of two sets of the multiple-photon transitions from the ground state in the QND-based spectrum of Fig.~2(b) of the main text. \textbf{(a)} Negative-detuning set, corresponding to the transitions $|g,0\rangle\to|n,-\rangle$. \textbf{(b)} Positive-detuning set, corresponding to the transitions $|g,0\rangle\to|n,+\rangle$. The horizontal and vertical axes are the drive frequency $\omega_d/2\pi$ and the drive power $P_d$, as in Fig.~2(b). Black arrows mark the individually resolved transition ridges; the marked sequence extends up to $n=20$ in (a) and $n=14$ in (b). The asymmetry between the upper and lower branches arises from the transmon anharmonicity. As $n$ increases, neighboring ridges crowd together and the whole pattern is displaced toward positive detuning.}
\label{fig:ground_sets_zoom}
\end{figure}

The practical consequence of separating the drive and readout is most evident when one counts how many multiphoton transitions from the ground state can still be distinguished, as compared with the earlier measurements discussed in the main text~\cite{bishopNonlinearResponseVacuum2009}. Figure~\ref{fig:ground_sets_zoom} shows enlarged views of the two ground-origin transition sets in Fig.~2(b) of the main text, in which the black arrows mark the ridges that remain individually discernible; the sequences extend up to $n=20$ for the negative-detuning set and $n=14$ for the positive-detuning set. As expected from $\omega_d^{g\to n,\pm}=\omega_c\pm g/\sqrt{n}$, neighboring ridges crowd together as $n$ increases and eventually broaden into a single feature, which sets the counting limit.

\subsection{High-resolution spectra and data processing}
\label{sec:data_processing}

\begin{figure}[htb]
\centering
\includegraphics[width=\columnwidth]{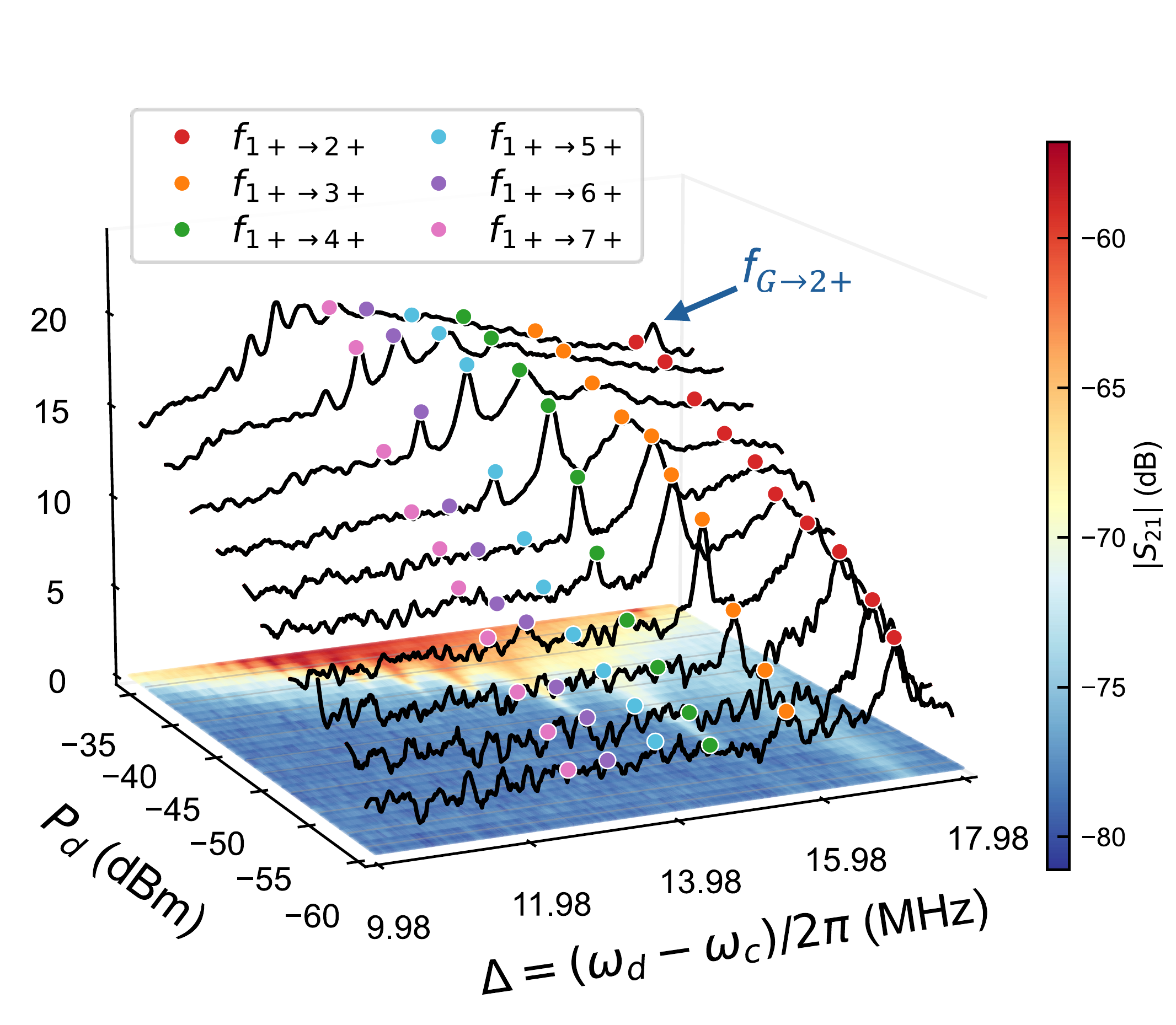}
\caption{Fine scan of the region outlined by the black box in Fig.~2(b) of the main text, measured by QND-based spectroscopy. The resolved ridges correspond to the thermally assisted $(n-1)$-photon transitions $|1,+\rangle\rightarrow|n,+\rangle$ and show their evolution with drive power. The overlaid black curves are representative smoothed traces used only to guide the eye.}
\label{fig:3D_plot}
\end{figure}

To demonstrate the enhanced spectral resolution enabled by QND-based spectroscopy and to illustrate the frequency-selection procedure used for Fig.~2(e) in the main text, we examine in Fig.~\ref{fig:3D_plot} a fine scan of the region outlined by the black box in Fig.~2(b) of the main text. The scan focuses on the higher-frequency set of thermally assisted transitions $|1,+\rangle\rightarrow|n,+\rangle$. The individual ridges remain well resolved as the drive power is varied, allowing their evolution to be followed clearly.

The measured two-dimensional spectrum is shown together with ten representative constant-power traces chosen at approximately equal power intervals. For the black overlays, each trace is smoothed along the frequency axis with a Savitzky--Golay filter using an $11$-point window and a second-order polynomial, rescaled to a common display span, and vertically displaced for clarity. These operations suppress point-to-point noise and make the power evolution of the resolved ridges easier to follow; they are used only for visualization.

For the quantitative comparison in Fig.~2(e), we instead used the unfiltered two-dimensional data. The experimental frequencies were selected manually at the centers of the resolved ridges; the smoothed traces in Fig.~\ref{fig:3D_plot} did not enter this procedure. Because several features are partially overlapping or asymmetric, we did not impose a common parametric line shape on all peaks. No separate statistical uncertainty was assigned to the manually selected points; the residuals below Fig.~2(e) quantify the experiment--model frequency differences rather than confidence intervals.

All $28$ selected frequencies were included with equal weight in a global transition-frequency fit to the numerically diagonalized multilevel model [Eq.~\eqref{eq:multiphoton_frequency}]. The fit jointly refined $\omega_c$, $g$, and $\Delta_{qc}$ and returned values consistent with the independently measured parameters of Table~\ref{tab:hamiltonian_parameters}: $\omega_c/2\pi=3.171~\mathrm{GHz}$, $g/2\pi=31~\mathrm{MHz}$, and $\Delta_{qc}=+0.47~\mathrm{MHz}$. The rms frequency residual was $0.16~\mathrm{MHz}$.


\section{Theoretical model}
\label{sec:model}

This section derives the multilevel eigenenergies and transition frequencies used in Figs.~2--4 of the main text. We first construct the multilevel eigenstates, then take the ideal two-level JC limit, and finally discuss the multilevel and residual-detuning corrections.

\subsection{Block-matrix method for calculating eigenvalues and transition frequencies}

We use the undriven multilevel Hamiltonian given in Eq.~(2) of the main text. For a complete description, we retain both finite qubit--cavity detuning $\Delta_{qc}=(\omega_q-\omega_c)/2\pi$ and the higher transmon levels. At each flux point, the qubit frequency is the calibrated value $\omega_q(\Phi)$. For simplicity, we denote this value by $\omega_q$ below. Within the rotating-wave approximation, the total excitation number $\hat N=\hat a^\dagger\hat a+\hat b^\dagger\hat b$ is conserved. The Hamiltonian is therefore block diagonal in subspaces with fixed excitation number $n$.

For $n\geq1$, let the transmon truncation contain $d_q$ levels and define $J_n=\min(n,d_q-1)$. The fixed-excitation basis is
\begin{equation}
\mathcal{B}_n
=
\bigl\{|j,n-j\rangle: j=0,\ldots,J_n\bigr\},
\label{eq:fixed_excitation_basis}
\end{equation}
where $|j,n-j\rangle=|j\rangle_{\rm q}\otimes|n-j\rangle_{\rm c}$ denotes the product state. The indices $j=0,1,2$ correspond to $|g\rangle$, $|e\rangle$, and $|f\rangle$, respectively. Restricting the Hamiltonian to this basis gives
\begin{equation}
\begin{aligned}
\frac{\hat H_n}{\hbar}
=&
\sum_{j=0}^{J_n}
\varepsilon_{jn}|j,n-j\rangle\langle j,n-j|
\\
&+
\sum_{j=0}^{J_n-1}
G_{jn}
\left(
|j+1,n-j-1\rangle\langle j,n-j|
+\mathrm{H.c.}
\right),
\\
\varepsilon_{jn}
={}&
(n-j)\omega_c+j\omega_q
-\frac{\alpha}{2}j(j-1),
\\
G_{jn}
={}&
g\sqrt{(n-j)(j+1)}.
\end{aligned}
\label{eq:fixed_excitation_hamiltonian}
\end{equation}
Here $\varepsilon_{jn}$ is the bare angular frequency of $|j,n-j\rangle$, and $G_{jn}=g\sqrt{(n-j)(j+1)}$ is the coupling to the adjacent basis state. The interaction changes the transmon and cavity excitation numbers in opposite directions, so no matrix elements connect nonadjacent values of $j$. We use $H_n$ for the matrix representation of $\hat H_n$ in $\mathcal B_n$. Each block is diagonalized according to
\begin{equation}
H_n|n,r\rangle=E_{n,r}|n,r\rangle,
\qquad
|n,r\rangle=\sum_{j=0}^{J_n}c_j^{(n,r)}|j,n-j\rangle.
\label{eq:fixed_excitation_eigenstates}
\end{equation}
Here $r$ labels the eigenstates within the $n$-excitation block, and $|n,r\rangle$ thus denotes the dressed states. The expansion coefficients satisfy the normalization condition $\sum_{j=0}^{J_n}|c_j^{(n,r)}|^2=1$. The ground state is $|g,0\rangle$ with $E_0=0$. The finite transmon and cavity cutoffs are chosen to include the excitation blocks needed for the spectral range shown below. For the one-excitation basis $\{|g,1\rangle,|e,0\rangle\}$,
\begin{equation}
\begin{aligned}
\frac{H_1}{\hbar}
&=
\begin{pmatrix}
\omega_c & g\\
g & \omega_q
\end{pmatrix},
\\
\frac{E_{1,s}}{\hbar}
&=
\frac{\omega_c+\omega_q}{2}+s\Omega_1,
\\
\Omega_1
&=\sqrt{g^2+\frac{(\omega_q-\omega_c)^2}{4}},
\end{aligned}
\label{eq:one_excitation_flux}
\end{equation}
where $s=\pm1$ and $\Omega_1$ is an angular frequency. At exact qubit--cavity resonance, this block gives the usual splitting $2g$.

The two-excitation block is needed for the multilevel transitions in Fig.~4. In the basis $\{|g,2\rangle,|e,1\rangle,|f,0\rangle\}$, subtracting the common energy $2\hbar\omega_c$ gives
\begin{equation}
\begin{aligned}
\widetilde H_2
&=H_2-2\hbar\omega_c I,
\\
\frac{\widetilde H_2}{\hbar}
&=
\begin{pmatrix}
0 & \sqrt{2}g & 0\\
\sqrt{2}g & 2\pi\Delta_{qc} & \sqrt{2}g\\
0 & \sqrt{2}g & 4\pi\Delta_{qc}-\alpha
\end{pmatrix}.
\end{aligned}
\label{eq:two_excitation_relative}
\end{equation}
The two nonzero off-diagonal entries describe the exchanges $|g,2\rangle\leftrightarrow|e,1\rangle$ and $|e,1\rangle\leftrightarrow|f,0\rangle$, each with matrix element $\sqrt{2}g$. The corner element between $|g,2\rangle$ and $|f,0\rangle$ vanishes because these states differ by two transmon excitations. Diagonalizing this matrix gives three flux-dependent eigenvalues $x_r(\Phi)$ defined by $E_{2,r}/h=2(\omega_c/2\pi)+x_r(\Phi)$ and the corresponding states $|2,r\rangle$. We label them $r=0,-,+$ by their dominant bare-state character away from the avoided crossings. Tracking these eigenvalues as a function of $\Phi$ produces the purple dressed-state curves in the $n=2$ excitation subspace shown in Fig.~4(a) of the main text.

For an $m$-photon transition from the initial state $|\psi_i\rangle$ to the final state $|\psi_f\rangle$, the resonance frequency is determined by
\begin{equation}
\omega_{if}^{(m)}
=
\frac{E_f-E_i}{m\hbar}.
\label{eq:multiphoton_frequency}
\end{equation}
The transition curves in Figs.~2(e) and 4(b) of the main text are obtained from Eq.~\eqref{eq:multiphoton_frequency}, with $|\psi_i\rangle=|g,0\rangle$ or $|1,\pm\rangle$ as appropriate. We first derive the near-resonant expressions in the ideal two-level JC limit and then include multilevel and residual-detuning corrections, using $\alpha/2\pi=192~\mathrm{MHz}$ from Table~\ref{tab:hamiltonian_parameters}.

\subsection{Ideal two-level JC limit}

The transition frequencies analyzed in Figs.~2 and 3 of the main text can be understood from the two-level limit of the multilevel Hamiltonian given in Eq.~(2) of the main text. We therefore retain only the $|g\rangle$ and $|e\rangle$ transmon levels. Using the fixed-excitation basis defined above, the $n\geq1$ block in the ordered basis $\{|g,n\rangle,|e,n-1\rangle\}$ at exact qubit--cavity resonance, $\omega_q=\omega_c$, is
\begin{equation}
\frac{\hat H_{n}}{\hbar}
=
n\omega_c\hat I
+g\sqrt{n}
\begin{pmatrix}
0&1\\
1&0
\end{pmatrix}.
\label{eq:jc_block}
\end{equation}
Solving the eigenvalue problem for this Hamiltonian gives the dressed states
\begin{equation}
|n,s\rangle
=
\frac{1}{\sqrt{2}}(|g,n\rangle+s|e,n-1\rangle),
\qquad s=\pm1,
\label{eq:ideal_jc_states}
\end{equation}
and the eigenenergies
\begin{equation}
\frac{E_{n,s}}{\hbar}
=
n\omega_c+s g\sqrt{n},
\qquad
s=\pm1,
\label{eq:ideal_jc_energies}
\end{equation}
where the energy of $|g,0\rangle$ is set to zero.

We first consider transitions originating from $|g,0\rangle$. Reaching $|n,s\rangle$ requires absorption of $n$ drive photons, so that $n\hbar\omega_d=E_{n,s}$. The resonance frequency is thus given by
\begin{equation}
\omega_d^{g\rightarrow n,s}
=
\omega_c+s\frac{g}{\sqrt{n}}.
\label{eq:ground_origin_jc}
\end{equation}
We next consider transitions originating from the thermally populated one-excitation dressed states. A transition $|1,s_i\rangle\rightarrow|n,s_f\rangle$ absorbs $n-1$ drive photons, and the corresponding energy difference is
$E_{n,s_f}-E_{1,s_i}=(n-1)\hbar\omega_c+\hbar g\left(s_f\sqrt{n}-s_i\right)$. Thus, we obtain the transition frequency
\begin{equation}
\omega_d^{s_i\rightarrow s_f}(n)
=
\omega_c
+g\frac{s_f\sqrt{n}-s_i}{n-1}.
\label{eq:thermal_jc_general}
\end{equation}
For both the ground-origin and thermally assisted processes, Eqs.~\eqref{eq:ground_origin_jc} and \eqref{eq:thermal_jc_general} give deviations from $\omega_c$ that scale as $n^{-1/2}$ at large $n$. Thus, as higher drive powers make transitions to progressively higher excitation subspaces accessible, all corresponding transition frequencies approach the bare-cavity frequency from above or below in the ideal two-level JC limit.

For the multiphoton transitions from the ground state plotted in Fig.~2(a) of the main text, the transition frequencies are
\begin{align}
|g,0\rangle&\rightarrow|1,\pm\rangle:
&\omega_d&=\omega_c\pm g,
\\
|g,0\rangle&\rightarrow|2,\pm\rangle:
&\omega_d&=\omega_c\pm\frac{g}{\sqrt{2}},
\\
|g,0\rangle&\rightarrow|3,\pm\rangle:
&\omega_d&=\omega_c\pm\frac{g}{\sqrt{3}},
\label{eq:ground_origin_low_order}
\end{align}
For transitions from the one-excitation dressed states, the transition frequencies for states of the same parity are
\begin{align}
\omega_d^{+\rightarrow+}(n)
&=
\omega_c+\frac{g}{\sqrt{n}+1},
\label{eq:thermal_jc_upper}
\\
\omega_d^{-\rightarrow-}(n)
&=
\omega_c-\frac{g}{\sqrt{n}+1},
\label{eq:thermal_jc_lower}
\end{align}
while those for states of different parity are
\begin{align}
\omega_d^{+\rightarrow-}(n)
&=
\omega_c-g\frac{\sqrt{n}+1}{n-1}
=
\omega_c-\frac{g}{\sqrt{n}-1},
\\
\omega_d^{-\rightarrow+}(n)
&=
\omega_c+g\frac{\sqrt{n}+1}{n-1}
=
\omega_c+\frac{g}{\sqrt{n}-1}.
\label{eq:thermal_jc_opposite}
\end{align}
In particular, the one-photon transitions resolved in Fig.~3 connect the one- and two-excitation dressed states:
\begin{align}
|1,+\rangle&\rightarrow|2,-\rangle:
&\omega_d&=\omega_c-(\sqrt{2}+1)g,
\\
|1,-\rangle&\rightarrow|2,+\rangle:
&\omega_d&=\omega_c+(\sqrt{2}+1)g.
\label{eq:opposite_jc_n2}
\end{align}

\subsection{Multilevel and residual-detuning corrections}

The ideal two-level JC model keeps only the bare states $|g,n\rangle$ and $|e,n-1\rangle$ in the fixed-excitation subspace with total excitation number $n$. It therefore does not reproduce the common upward frequency shift observed at large $n$. For a multilevel transmon and $n\geq2$, the same subspace also contains the bare state $|f,n-2\rangle$. This state couples directly to $|e,n-1\rangle$ through the cavity--transmon interaction and consequently hybridizes with both JC dressed states. The resulting level repulsion shifts both $|n,-\rangle$ and $|n,+\rangle$ upward. Because $|f,n-2\rangle$ is closer in energy to $|n,-\rangle$, the transition ending in $|n,-\rangle$ receives the larger shift. The numerical diagonalization includes this state and all higher transmon levels. We isolate it below to identify the leading physical correction.

To make this perturbative argument explicit, we use the resonant decomposition of the same three-state structure introduced in Eq.~\eqref{eq:two_excitation_relative}. This decomposition separates the ideal JC doublet from the residual coupling to $|f,n-2\rangle$ and is used to calculate the leading second-order multilevel correction. We restrict the $n\geq2$ fixed-excitation subspace to the basis $\{|g,n\rangle,|e,n-1\rangle,|f,n-2\rangle\}$ and write
\begin{equation}
\hat H_n=\hat H_n^{(0)}+\hat V_n,
\label{eq:multilevel_perturbative_split}
\end{equation}
where $\hat H_n^{(0)}$ contains the ideal JC doublet and the uncoupled $|f,n-2\rangle$ state, while $\hat V_n$ contains their residual coupling. In the stated basis and at exact qubit--cavity resonance, let $I_3$ denote the three-dimensional identity. The two matrix representations are
\begin{align}
\frac{\hat H_n^{(0)}}{\hbar}
&=n\omega_c I_3+
\begin{pmatrix}
0 & g\sqrt{n} & 0\\
g\sqrt{n} & 0 & 0\\
0 & 0 & -\alpha
\end{pmatrix},
\\
\frac{\hat V_n}{\hbar}
&=g\sqrt{2(n-1)}
\begin{pmatrix}
0 & 0 & 0\\
0 & 0 & 1\\
0 & 1 & 0
\end{pmatrix}.
\label{eq:multilevel_perturbative_parts}
\end{align}

The first matrix contains the common bare energy $n\omega_c$, the ideal JC coupling $g\sqrt{n}$ between $|g,n\rangle$ and $|e,n-1\rangle$, and the $-\alpha$ offset of $|f,n-2\rangle$. The perturbation $\hat V_n$ couples only $|e,n-1\rangle$ to $|f,n-2\rangle$ with matrix element $g\sqrt{2(n-1)}$; the direct $|g,n\rangle$--$|f,n-2\rangle$ matrix element is zero.

The unperturbed eigenstates in the JC doublet are $|n,s\rangle$ with $E_{n,s}^{(0)}=\hbar(n\omega_c+s g\sqrt{n})$, while $E_{f,n-2}^{(0)}=\hbar(n\omega_c-\alpha)$. Since $\hat V_n$ has no diagonal matrix element in $|n,s\rangle$, the first-order energy correction vanishes. The leading correction is therefore the second-order term
\begin{equation}
\delta E_{n,s}^{(2)}
=
\frac{\left|\langle f,n-2|\hat V_n|n,s\rangle\right|^2}
{E_{n,s}^{(0)}-E_{f,n-2}^{(0)}}.
\label{eq:multilevel_second_order_general}
\end{equation}
Using $|n,s\rangle=(|g,n\rangle+s|e,n-1\rangle)/\sqrt{2}$ in Eq.~\eqref{eq:multilevel_perturbative_parts} gives
\begin{align}
\langle f,n-2|\hat V_n|n,s\rangle=s\hbar g\sqrt{n-1},\\
E_{n,s}^{(0)}-E_{f,n-2}^{(0)}=\hbar(\alpha+s g\sqrt{n}).
\end{align}
Substitution into Eq.~\eqref{eq:multilevel_second_order_general} then yields
\begin{align}
\frac{\delta E_{n,+}}{\hbar}
&\simeq
\frac{(n-1)g^2}{\alpha+g\sqrt{n}},
\label{eq:multilevel_shift_upper}
\\
\frac{\delta E_{n,-}}{\hbar}
&\simeq
\frac{(n-1)g^2}{\alpha-g\sqrt{n}}.
\label{eq:multilevel_shift_lower}
\end{align}
The one-excitation subspace contains no $|f\rangle$ state, so $E_{1,\pm}$ is unchanged at this order. For a ground-origin $n$-photon transition, the final-state correction enters the drive frequency as $\delta E_{n,s}^{(2)}/(n\hbar)$, giving
\begin{equation}
\begin{aligned}
\omega_d^{g\rightarrow n,s}
&\simeq
\omega_c+s\frac{g}{\sqrt{n}}
\\
&\quad
+\frac{n-1}{n}
\frac{g^2}{\alpha+s g\sqrt{n}},
\qquad s=\pm1.
\end{aligned}
\label{eq:multilevel_ground_origin}
\end{equation}
For transitions originating from the one-excitation dressed states, the correction enters as $\delta E_{n,s}^{(2)}/[(n-1)\hbar]$, giving
\begin{align}
\omega_d^{+\rightarrow+}(n)
&\simeq
\omega_c+\frac{g}{\sqrt{n}+1}
+\frac{g^2}{\alpha+g\sqrt{n}},
\label{eq:multilevel_thermal_upper}
\\
\omega_d^{-\rightarrow-}(n)
&\simeq
\omega_c-\frac{g}{\sqrt{n}+1}
+\frac{g^2}{\alpha-g\sqrt{n}}.
\label{eq:multilevel_thermal_lower}
\end{align}
Representative detunings from these expressions are listed in Table~\ref{tab:multilevel_frequency_examples}. For this table, the residual qubit--cavity detuning $\Delta_{qc}\equiv(\omega_q-\omega_c)/2\pi$ is set to zero to isolate the multilevel correction.

\begin{table}[tb]
\caption{Representative drive detunings $(\omega_d-\omega_c)/2\pi$ in megahertz, evaluated using $g/2\pi=31~\mathrm{MHz}$ and $\alpha/2\pi=192~\mathrm{MHz}$. ``JC'' denotes the ideal two-level result, and ``ML'' includes the second-order correction in Eq.~\eqref{eq:multilevel_ground_origin} or Eqs.~\eqref{eq:multilevel_thermal_upper} and \eqref{eq:multilevel_thermal_lower}.}
\label{tab:multilevel_frequency_examples}
\centering
\begin{ruledtabular}
\begin{tabular}{ccccc}
\multicolumn{5}{c}{$|g,0\rangle\rightarrow|n,\pm\rangle$} \\
& \multicolumn{2}{c}{$|n,+\rangle$} & \multicolumn{2}{c}{$|n,-\rangle$} \\
$n$ & JC & ML & JC & ML \\
2 & $+21.92$ & $+23.96$ & $-21.92$ & $-18.68$ \\
4 & $+15.50$ & $+18.34$ & $-15.50$ & $-9.96$ \\
8 & $+10.96$ & $+13.97$ & $-10.96$ & $-2.90$ \\
\colrule
\multicolumn{5}{c}{$|1,\pm\rangle\rightarrow|n,\pm\rangle$} \\
& \multicolumn{2}{c}{$+\rightarrow+$} & \multicolumn{2}{c}{$-\rightarrow-$} \\
$n$ & JC & ML & JC & ML \\
2 & $+12.84$ & $+16.92$ & $-12.84$ & $-6.35$ \\
4 & $+10.33$ & $+14.12$ & $-10.33$ & $-2.94$ \\
8 & $+8.10$ & $+11.53$ & $-8.10$ & $+1.11$ \\
\end{tabular}
\end{ruledtabular}
\end{table}

For $g\sqrt{n}/\alpha\ll1$, the multilevel corrections to both the ground-origin and thermally assisted transition frequencies are positive and are larger for transitions ending in the lower dressed state, consistent with the level-repulsion picture above. Since this perturbative condition is no longer well satisfied at $n=8$, the corresponding table entries illustrate only the direction of the shift.

The residual qubit--cavity detuning $\Delta_{qc}=(\omega_q-\omega_c)/2\pi$ modifies the two-level dressed energies, in ordinary-frequency units, as
\begin{equation}
\frac{E_{n,s}}{h}
=
n\frac{\omega_c}{2\pi}+\frac{\Delta_{qc}}{2}
+s\sqrt{n\left(\frac{g}{2\pi}\right)^2
+\frac{\Delta_{qc}^2}{4}}.
\label{eq:detuned_jc_energies}
\end{equation}
For transitions between excited states, the common term $\Delta_{qc}/2$ cancels from the energy difference. The shared upward displacement of the thermally assisted transition sets is therefore predominantly a multilevel effect, rather than a correction linear in $\Delta_{qc}$. All curves in Fig.~2(e) of the main text are obtained by exact diagonalization of the multilevel Hamiltonian using the single globally fitted value $\Delta_{qc}\simeq+0.47~\mathrm{MHz}$.

\section{Numerical analysis}
\label{sec:numerics}

This section describes the numerical solution used to calculate the driven steady-state spectra and the observables compared with experiment. We first introduce the steady-state solver and its implementation, and then discuss the effective thermal excitation included in the model. Finally, we compare the calculated observables with the experimental spectra.

\subsection{Numerical solution of the master equation}
\label{sec:numerical_solution}

We use the Hamiltonian from the main text after transforming to the frame rotating at the drive frequency $\omega_d$. Then, the resulting master equation can be solved directly with QuTiP. We therefore do not repeat the Hamiltonian here. We use the ordinary-frequency detunings $\Delta=(\omega_d-\omega_c)/2\pi$ and $\Delta_{qc}=(\omega_q-\omega_c)/2\pi$. For each pair of drive frequency and drive amplitude, we solve the corresponding Liouvillian for the normalized steady state,
\begin{equation}
\mathcal{L}(\omega_d,\epsilon_d)\hat{\rho}_{\rm ss}=0,
\qquad
\operatorname{Tr}\hat{\rho}_{\rm ss}=1.
\label{eq:steady_state}
\end{equation}

The numerical procedure is a direct parameter sweep. We truncate the cavity and transmon Hilbert spaces, construct the tensor-product operators, build the rotating-frame Hamiltonian operator for each $\omega_d$, and use a steady-state Lindblad solver to obtain $\hat\rho_{\rm ss}$. Figure~\ref{fig:qutip_code} shows the core QuTiP implementation. The code first constructs the system operators (lines 4--8) and then defines the collapse operators associated with dissipation (lines 9--14). It next sweeps the drive power and frequency and solves for the steady state (lines 15--27), and then evaluates the two simulated responses (lines 28--33). All frequencies are expressed in one consistent unit and $\hbar=1$.

\clearpage % flush pending floats so the tall code listing can be placed
\begin{figure}[tb]
\centering
\begin{lstlisting}[style=qutipcode, firstnumber=1,
    basicstyle=\ttfamily\fontsize{9.4}{9}\selectfont]
import numpy as np
import qutip as qt
Nc, Nq = 8, 4
a = qt.tensor(qt.destroy(Nc),
              qt.qeye(Nq))
b = qt.tensor(qt.qeye(Nc),
              qt.destroy(Nq))
nb = b.dag() * b
c_ops = [
    np.sqrt(kappa * (n_th + 1)) * a,
    np.sqrt(kappa * n_th) * a.dag(),
    np.sqrt(gamma_1) * b,
    np.sqrt(gamma_phi) * b.dag() * b,
]
for Pd in drive_powers:
    eps = epsilon_ref * 10**(
        (Pd - P_ref) / 20)
    for wd in drive_frequencies:
        H = (
            (omega_c - wd) * a.dag() * a
            + (omega_q - wd) * b.dag() * b
            - alpha / 2 * b.dag()**2 * b**2
            + g * (a.dag() * b
                   + a * b.dag())
            + eps * (a + a.dag())
        )
        rho_ss = qt.steadystate(H, c_ops)
        nq = qt.expect(nb, rho_ss)
        ac = qt.expect(a, rho_ss)
        Yq = 20 * np.log10(
            max(nq, 1e-300))
        Yc = 10 * np.log10(
            max(abs(ac / eps)**2, 1e-300))
\end{lstlisting}
\caption{Core QuTiP implementation of the steady-state simulation. The code constructs the truncated cavity--transmon Hilbert space and collapse operators, sweeps the drive frequency $\omega_d$ and power $P_d$, solves the Lindblad steady state, and evaluates the transmon-excitation response $Y_q$ and normalized coherent cavity response $Y_c$.}
\label{fig:qutip_code}
\end{figure}

The code uses the transmon coefficients $\gamma_1$ and $\gamma_\varphi$. With the dissipator convention used in the main text, $\gamma_1\mathcal{D}[\hat b]$ gives the population-relaxation rate, while $\gamma_\varphi\mathcal{D}[\hat b^\dagger\hat b]$ contributes $\gamma_\varphi/2$ to the Ramsey pure-dephasing rate. Thus,
\begin{equation}
\gamma_1=\frac{1}{T_1},
\qquad
\gamma_\varphi=\frac{2}{T_2^*}-\gamma_1.
\label{eq:transmon_rate_conversion}
\end{equation}
For $T_1=38~\mu\mathrm{s}$ and $T_2^*=4.6~\mu\mathrm{s}$, we obtain $\gamma_1=0.0263~\mu\mathrm{s}^{-1}$ and $\gamma_\varphi=0.4085~\mu\mathrm{s}^{-1}$. 

The QND readout is used as a proxy for the steady-state transmon excitation. We therefore evaluate
\begin{equation}
n_q^{\rm ss}
=
\langle\hat b^\dagger\hat b\rangle_{\rm ss}
=
\operatorname{Tr}\!\left(\hat b^\dagger\hat b\hat{\rho}_{\rm ss}\right).
\label{eq:transmon_excitation}
\end{equation}
We store $Y_q=20\log_{10}n_q^{\rm ss}$ in the numerical sweep.
In conventional cavity spectroscopy, the transmitted coherent power is proportional to $|\langle\hat a\rangle_{\rm ss}|^2$, rather than to the mean intracavity occupation $\langle\hat a^\dagger\hat a\rangle_{\rm ss}$. We therefore use the normalized coherent intracavity response
\begin{equation}
Y_c
=
10\log_{10}
\left|
\frac{\langle\hat a\rangle_{\rm ss}}{\epsilon_d}
\right|^2,
\label{eq:cavity_response}
\end{equation}
up to an arbitrary constant offset.

\subsection{Effective thermal occupation of the main cavity}
\label{sec:thermal_excitation}

The effective thermal occupation of the main cavity is parameterized as
\begin{equation}
\nth
=
\left[
\exp\!\left(\frac{\hbar\omega_c}{k_{\rm B}T_{\rm eff}}\right)-1
\right]^{-1}.
\label{eq:thermal_occupation}
\end{equation}
Here $T_{\rm eff}$ is a convenient parameterization of the fitted bath occupation, not an independently measured equilibrium temperature of the driven device. A single value of $\nth$ is used for the calculations in the main text and in the supplemental steady-state spectra. In this effective model, the finite upward transition rate is assigned to the main-cavity bath. Near resonance, cavity--transmon hybridization transfers this bath occupation to the dressed states and thereby populates the initial states of the thermally assisted transitions.

A cavity bath and a transmon bath act through different operators, $\hat a$ and $\hat b$, and therefore generate different dressed-state transition rates and steady-state populations. As a model check, we assigned the finite thermal occupation exclusively to the transmon bath while keeping the cavity bath at zero occupation. The resulting spectrum differed qualitatively from the experiment. We therefore retain the cavity-bath term as the minimal effective representation of the residual environmental excitation supported by the measured spectra. This choice does not imply that the physical cryogenic environment couples exclusively to the cavity.

\subsection{Numerical simulations of the experimental results}
\label{sec:numerical_simulations}

\begin{figure*}[t]
\centering
{\includegraphics[width=0.96\linewidth]{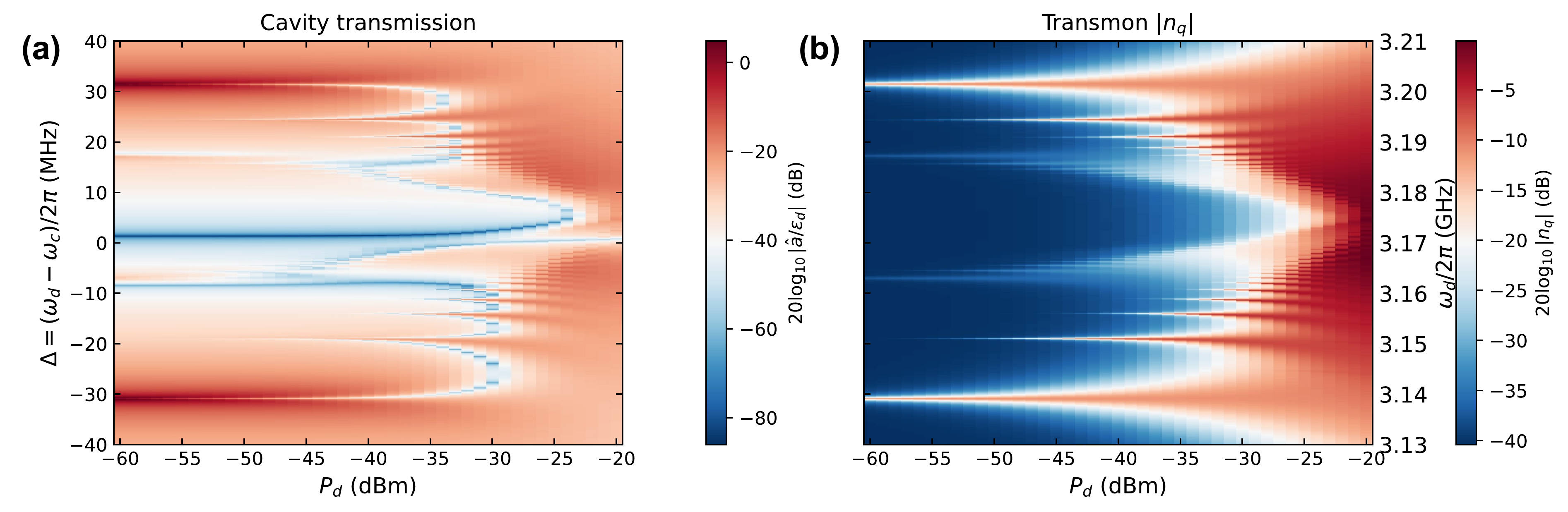}}
\caption{Comparison of the calculated responses with the measured QND spectrum in Fig.~2(b) of the main text. The drive-detuning and drive-power axes use the same definitions, enabling a direct comparison. \textbf{(a)} Normalized coherent intracavity response, $Y_c=10\log_{10}|\langle\hat a\rangle_{\rm ss}/\epsilon_d|^2$, used as a proxy for direct cavity transmission. \textbf{(b)} Mean transmon-excitation response, $Y_q=20\log_{10}n_q^{\rm ss}$, which is the calculated QND-readout counterpart of Fig.~2(b). Both maps are obtained from the same steady-state density matrices using the globally fitted $\nth$ and the fixed drive calibration in Eq.~\eqref{eq:drive_conversion}, with no power-dependent refitting. The calculated transition lines agree with the measured transition positions and their evolution with drive power.}
\label{fig:large_range_simulation}
\end{figure*}

\begin{table}[htb]
\caption{Effective model fit parameters used for the finite-occupation steady-state calculation in Fig.~2(d) of the main text. The independently characterized low-power device values are listed in Table~\ref{tab:hamiltonian_parameters}. The transmon rates are fixed model inputs rather than fit parameters. The dimensions $d_c$ and $d_q$ denote the retained cavity Fock states and transmon levels, respectively.}
\label{tab:steady_state_parameters}
\centering
\begin{ruledtabular}
\begin{tabular}{lcc}
Quantity & Symbol & Value \\
Effective cavity-bath occupation & $\nth$ & $0.010$ \\
Fitted cavity decay rate & $\kappa/2\pi$ & $0.0611~\mathrm{MHz}$ \\
Reference drive amplitude (relative) & $\epsilon_{\rm ref}/2\pi$ & $0.1108~\mathrm{MHz}$ \\
Hilbert-space dimensions & $d_c\times d_q$ & $8\times4$ \\
\end{tabular}
\end{ruledtabular}
\end{table}

With the thermal model specified, we use the steady-state solutions to simulate the measured line cuts and power-dependent spectra. The value of $\nth$ and the conversion from the reported drive power $P_d$ to the Hamiltonian drive amplitude are obtained from a joint steady-state fit to the three line cuts in Fig.~2(d) of the main text at $P_d=-60$, $-50$, and $-40~\mathrm{dBm}$. The global physical fit parameters are $\nth$, $\kappa$, and the reference drive amplitude $\epsilon_{\rm ref}$. The Hamiltonian parameters and transmon rates are fixed, and each trace has one additive vertical offset. The drive at any other power is constrained by
\begin{equation}
\epsilon_d(P_d)
=
\epsilon_{\rm ref}
10^{(P_d-P_{\rm ref})/20},
\qquad
P_{\rm ref}=-60~\mathrm{dBm}.
\label{eq:drive_conversion}
\end{equation}

The reduced model does not include the separate readout resonator. It therefore
reproduces the spectral structure of the response but not its absolute level, and
each calculated trace is compared with the measured one up to one additive vertical offset. Because such an offset is degenerate with a common rescaling of the drive amplitude, $\epsilon_{\rm ref}$ fixes only the relative drive scaling between traces; it is not an absolutely calibrated amplitude and is not converted into an intracavity photon number. We compare only the spectral structure, using $Y_q=20\log_{10}n_q^{\rm ss}$ as a phenomenological logarithmic proxy and allowing one constant vertical offset for each power in the line-cut fit. Neither $Y_q$ nor the coherent response $Y_c$ in Eq.~\eqref{eq:cavity_response} is interpreted as an absolutely calibrated transmission. The remaining residuals from the three traces are weighted equally in the joint fit.

The parameter set used for the displayed steady-state curves is summarized in Table~\ref{tab:steady_state_parameters}: $\nth=0.010$, $\kappa/2\pi=0.0611~\mathrm{MHz}$, and $\epsilon_{\rm ref}/2\pi=0.1108~\mathrm{MHz}$. Equation~\eqref{eq:drive_conversion} then gives $\epsilon_d/2\pi=0.3505$ and $1.1083~\mathrm{MHz}$ at $-50$ and $-40~\mathrm{dBm}$, respectively. The joint rms discrepancy in the logarithmic response is $1.68~\mathrm{dB}$. These values parameterize the displayed model comparison; no formal parameter uncertainties are estimated or reported from this fit. The same effective $\nth$ is subsequently fixed in the calculation for Fig.~3 of the main text. For the zero-occupation case, only $\nth$ is set to zero; the truncation and all other parameters are unchanged.

To test the model beyond the selected line cuts, we
also solve Eq.~\eqref{eq:steady_state} on a grid spanning $-40\leq\Delta\leq40~\mathrm{MHz}$ and $-60\leq P_d\leq-20~\mathrm{dBm}$ as shown in Fig.~\ref{fig:large_range_simulation}. It is constructed on the same drive-detuning and drive-power coordinates as the measured spectrum in Fig.~2(b) of the main text, so that it can be compared directly with that panel. In particular, panel (b) is the calculated QND-readout counterpart of Fig.~2(b), while panel (a) shows the conventional-transmission response obtained from the same steady states.
At each grid point, we obtain $Y_c$ for conventional transmission spectroscopy and $Y_q$ for QND-based spectroscopy from the same steady-state density matrix. We use the same globally fitted $\nth$, dissipative parameters, and fixed power-to-drive conversion as in the line-cut comparison, without any power-dependent refitting. The simulated positions and power evolution of the resolved ground-origin and thermally assisted transition sets remain consistent with the measured spectra, providing a check of the driven-dissipative description beyond the fitted line cuts.

At high drive powers, the experimental spectra additionally develop a bare-cavity response at $\Delta=0$, arising from the bright component of the bistable photon-blockade-breakdown response \cite{finkObservationPhotonBlockadeBreakdown2017,settEmergentMacroscopicBistability2024}. The present finite-truncation calculation does not reproduce this high-occupation central component. We therefore do not use the simulated response in this region to interpret the central feature; its absence does not enter the finite-detuning transition assignments.

For this large-range sweep, four transmon levels are retained and the cavity cutoff is increased with drive power from $4$ to $20$ Fock states. The maximum calculated cavity occupation is monitored relative to the cutoff for each power row to identify any region that requires a larger cavity basis.

The additional comparisons in Figs.~3 and 4 of the main text are obtained using the calibrated device parameters and model conventions described above, without introducing new fit parameters. For Fig.~3, we use the same steady-state model and parameters described above. For Fig.~4, we use the calibrated device parameters to calculate the flux-dependent transition frequencies. The resulting spectrum and transition curves agree well with the corresponding measurements, and we do not repeat these calculations here.

\section{Flux-resolved transitions and virtual two-photon exchange}
\label{sec:flux_transitions}

This section connects the flux-dependent eigenenergies obtained above to the transition curves in Fig.~4 of the main text. We first identify the three transition sets shown in Fig.~4(b). We then derive the virtual-photon-mediated coupling at the two-excitation avoided crossing. Finally, we compare how the same energy-level splitting appears in one- and two-photon spectra.

We use the one- and two-excitation eigenenergies and the general relation in Eq.~\eqref{eq:multiphoton_frequency}; no new eigenvalue problem is introduced here. The black curves are the ground-origin one-photon transitions $|g,0\rangle\rightarrow|1,s\rangle$, with frequencies
\begin{equation}
\begin{aligned}
\frac{\omega_{d}^{g0\rightarrow 1,s}(\Phi)}{2\pi}
&=
\frac{E_{1,s}(\Phi)-E_0}{h}
\\
&=
\frac{\omega_c}{2\pi}
+\frac{\Delta_{qc}(\Phi)}{2}
+s\frac{\Omega_1(\Phi)}{2\pi},
\end{aligned}
\label{eq:fig4_ground_one_photon}
\end{equation}
where $s=\pm1$. At $\Delta_{qc}=0$, the coupling $g$ hybridizes $|g,1\rangle$ and $|e,0\rangle$, producing the black avoided crossing with an angular-frequency splitting $2g$.

The yellow curves are the ground-origin two-photon transitions $|g,0\rangle\rightarrow|2,r\rangle$, with frequencies
\begin{equation}
\begin{aligned}
\frac{\omega_{d}^{g0\rightarrow 2,r}(\Phi)}{2\pi}
&=
\frac{E_{2,r}(\Phi)-E_0}{2h}
\\
&=
\frac{\omega_c}{2\pi}+\frac{x_r(\Phi)}{2},
\end{aligned}
\label{eq:fig4_ground_two_photon}
\end{equation}
where $r=0,-,+$. The distinctive weak avoided crossing in Fig.~4 occurs at $\Delta_{qc}=\alpha/(4\pi)$, where $|g,2\rangle$ and $|f,0\rangle$ are degenerate. These states have no direct matrix element; the coupling is mediated by the off-resonant intermediate state $|e,1\rangle$. The direct-exchange crossings are already contained in the fixed-excitation matrix in Eq.~\eqref{eq:two_excitation_relative}.

At $\Delta_{qc}=\alpha/(4\pi)$, the diagonal angular frequencies in Eq.~\eqref{eq:two_excitation_relative} are $0$, $\alpha/2$, and $0$. The intermediate state $|e,1\rangle$ is therefore detuned from the degenerate pair by $\alpha/2$, and its coupling to each state is $\sqrt{2}g$. Direct diagonalization of the three-state matrix gives
\begin{equation}
\frac{\Delta E_{\mathrm{gap}}^{(3\text{-state})}}{\hbar}
=
\sqrt{\left(\frac{\alpha}{4}\right)^2+4g^2}
-\frac{\alpha}{4},
\label{eq:two_photon_gap_exact}
\end{equation}
which is the energy-level gap divided by $\hbar$; the corresponding splitting in the plotted ordinary-frequency axis is $\Delta\omega_{\rm gap}/2\pi$. This virtual two-photon exchange is an internal second-order interaction within the undriven system and should not be confused with the absorption of two drive photons in Eq.~\eqref{eq:fig4_ground_two_photon}.

The green curves arise from the six one-photon transitions $|1,s\rangle\rightarrow|2,r\rangle$ enabled by the finite population of the one-excitation dressed states. Their frequencies are
\begin{align}
\frac{\omega_{d}^{1,s\rightarrow 2,r}(\Phi)}{2\pi}
&=
\frac{E_{2,r}(\Phi)-E_{1,s}(\Phi)}{h}
\nonumber\\
&=
\frac{\omega_c}{2\pi}+x_r(\Phi)
-\frac{\Delta_{qc}(\Phi)}{2}
-s\frac{\Omega_1(\Phi)}{2\pi}.
\label{eq:fig4_dressed_origin}
\end{align}
Near $\Delta_{qc}=0$, the $\pm\rightarrow\pm$ and $\pm\rightarrow\mp$ subsets give, respectively, the nearly parallel same-parity transition curves and the oppositely oriented parabolic opposite-parity transition curves described in the main text. For two transitions sharing the same initial state, the initial-state energy cancels, giving
\begin{equation}
\omega_{d}^{1,s\rightarrow2,r'}
-\omega_{d}^{1,s\rightarrow2,r}
=
\frac{E_{2,r'}-E_{2,r}}{\hbar}.
\label{eq:fig4_common_initial_splitting}
\end{equation}
Near $\Delta_{qc}=\alpha/(4\pi)$, the green transition from $|1,-\rangle$ and the yellow transition from $|g,0\rangle$ terminate in the same pair of states, $|2,-\rangle$ and $|2,0\rangle$. Although they start from different initial states and involve different numbers of drive photons, they probe the same energy-level gap, which we denote by $\Delta\omega_{\rm gap}=(E_{2,0}-E_{2,-})/\hbar$. For the green one-photon transitions, the initial-state energy cancels and the separation of the drive frequencies is $\Delta\omega_{\rm gap}$. For the yellow two-photon transitions, the same final-state energy difference is divided by two drive photons, so their drive-frequency separation is $\Delta\omega_{\rm gap}/2$. Thus, the same avoided crossing appears with a factor-of-two difference in the transition-frequency splitting.

\begin{figure*}[tb]
\centering
\includegraphics[width=\linewidth]{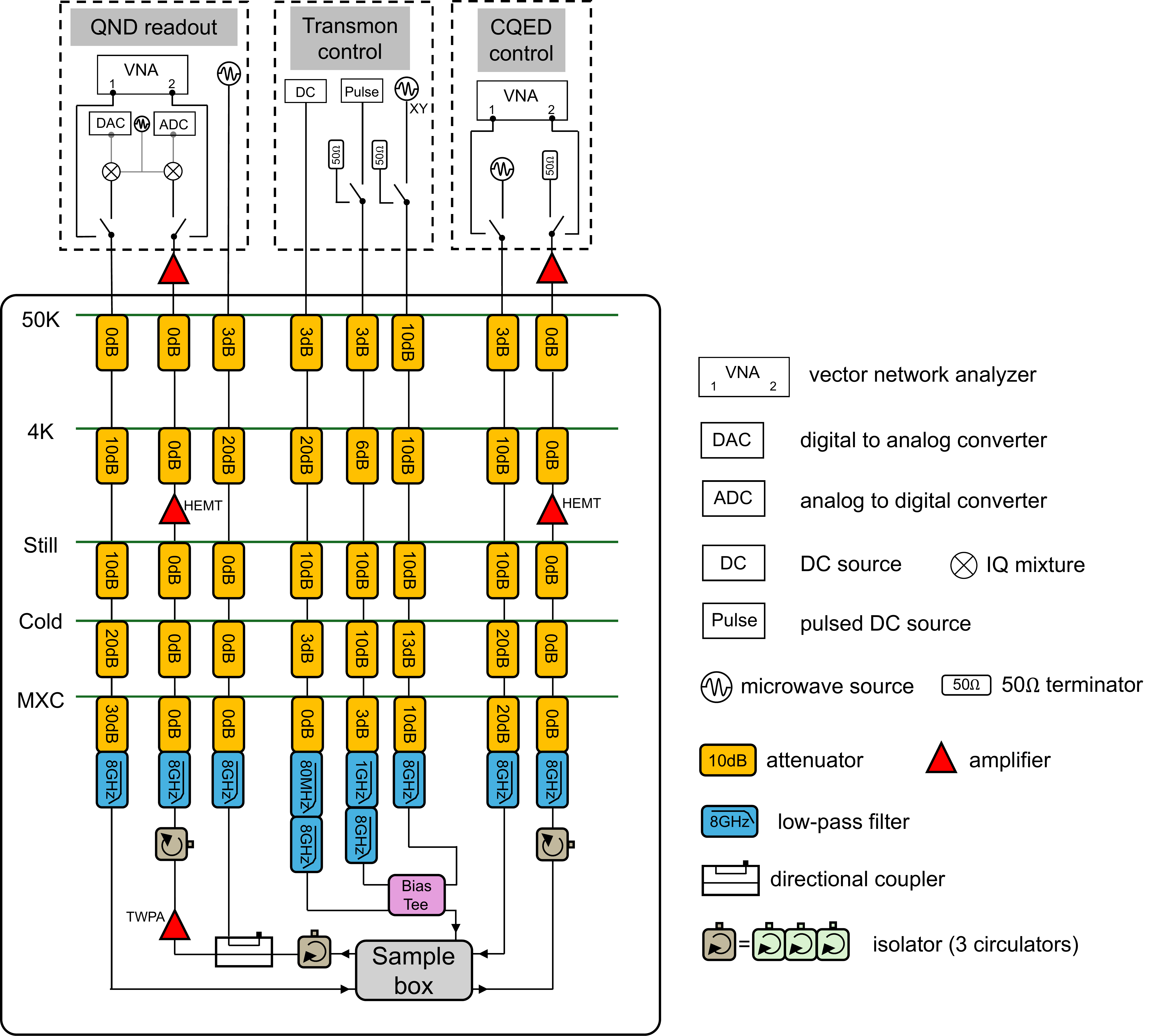}
\caption{Complete cryogenic microwave wiring and room-temperature instrumentation. The spectroscopy drive and QND readout use separate microwave lines. The readout instrumentation includes the pulsed DAC/ADC configuration used together with the transmon-control line and the VNA-based CW configuration. Attenuators, filters, isolating components, and amplifiers are shown at their corresponding temperature stages. Note that, in addition to the attenuation indicated in the figure, the stainless-steel semi-rigid coaxial cables and connectors between the attenuators also introduce additional loss. From separate measurements, we estimate this excess attenuation to be about 22 dB along a single control line from room temperature to the mixing chamber. For the readout lines, which include the HEMT amplifier, the excess attenuation is about 13 dB; because the Al and NbTi coaxial cables used in the mixing-chamber and cold-still stages are superconducting, this loss originates mainly from the connectors rather than from the cable dielectric.}
\label{fig:setup}
\end{figure*}

For reference, the complete cryogenic wiring and room-temperature instrumentation used throughout the experiment are shown in Fig.~\ref{fig:setup}.

\end{document}